\documentclass[12pt]{JHEP3}
\usepackage{amsmath}
\usepackage{epsfig}
\usepackage{amssymb,amsfonts}
\usepackage[all]{xy}

\def\Tr{\,{\rm Tr}\, }

\def\IP{\relax{\rm I\kern-.18em P}}

\def\be{\begin{equation}}
\def\ee{\end{equation}}
\def\ba{\begin{eqnarray}}
\def\ea{\end{eqnarray}}
\newcommand{\A}{{\cal A}}

\renewcommand{\H}{{\cal H}}

\newcommand{\Zop}{\mathbb{Z}}

\newcommand{\half}{\frac{1}{2}}

\def\BC{{\mathbb C}}
\def\lto{\longrightarrow}
\def\Coker{\operatorname{Coker}\,}
\def\beq{\begin{equation}}
\def\eeq{\end{equation}}
\def\beqa{\begin{eqnarray}}
\def\eeqa{\end{eqnarray}}
\newcommand{\ZZ}{\mathbb{Z}}

\def\sqr#1#2{{
\vcenter{\vbox{\hrule height.#2pt
\hbox{\vrule width.#2pt height#1pt \kern#1pt
\vrule width.#2pt}
\hrule height.#2pt}}}}
\boldmath
\title{Matrix factorisations and permutation branes}
\unboldmath
\author{Ilka Brunner and Matthias R.\ Gaberdiel
\\
$\ $ \\
$\ $ \\
Institut f\"ur Theoretische Physik, ETH-H\"onggerberg\\
8093 Z\"urich, Switzerland\\
Email: \email{brunner@itp.phys.ethz.ch}, 
\email{gaberdiel@itp.phys.ethz.ch}
}

\abstract{The description of B-type D-branes on a tensor
product of two $N=2$ minimal models in terms of matrix factorisations
is related to the boundary state description in 
conformal field theory. As an application we show that the D0- and
D2-brane for a number of Gepner models are described by permutation
boundary states. In some cases (including the quintic) the images of
the D2-brane under the Gepner monodromy generate the full charge
lattice.}

\preprint{hep-th/0503207}


\begin{document}

\setcounter{section}{0}
\section{Introduction}

Recently, Maxim Kontsevich has suggested that supersymmetric B-type
D-branes in Landau-Ginzburg models can be characterised in terms of 
matrix factorisations
\be
Q^2 = W \cdot {\bf 1} \,,
\ee
where $W(\Phi)$ is the Landau-Ginzburg superpotential of the
superfield $\Phi$. Here $Q$ is an off-diagonal matrix 
\be
Q = \begin{pmatrix}
 0 & J \cr E & 0 
\end{pmatrix} \,,
\ee
and $E$ and $J$ are polynomial matrices in the superfield $\Phi$. The
matrices $E$ and $J$ appear in the action in the boundary F-terms,
that also involve fermionic fields living at the boundary.
Their presence
is required in order to cancel a  boundary term in the supersymmetry
variation of the bulk F-term 
\be
\int_{\Sigma} d^2x\, d\theta^+ d\theta^- \, W + {\rm c.c.} \,.
\ee
This approach was proposed in unpublished form by Kontsevich, and the 
physical interpretation of it was given in \cite{KL1,BHLS,KL2,KL3,L,HL}; 
for a good review of this material see for example \cite{HWs}.

{}From a space-time point of view, these world-sheet fermions describe
open string tachyons that appear in brane anti-brane configurations. 
[Recall that the open string GSO-projection for an open string between
a brane and an anti-brane is opposite to that for the open string
between two branes or two anti-branes; the above world-sheet fermions
are projected out between branes and branes (or anti-branes and
anti-branes), but survive the GSO-projection for open strings between
branes and anti-branes.] Thus the matrix $Q$ can also be thought of as
describing the tachyonic configuration on (spacetime filling) brane
anti-brane pairs that leads to the D-brane in question. 
\smallskip

On the other hand, at least certain Landau-Ginzburg models have a
microscopic description in terms of $N=2$ minimal models  
\cite{Vafa1,Mar1,VW,W,Gep}. These D-branes must therefore also have a
conformal field theoretic description. It is then interesting to 
understand the relation between these two points of view in detail. 

For the case of a single minimal model, this correspondence has been
understood \cite{BHLS,KL2,KL3}, but in general little is known. In 
this paper we want to study the next simple case,  
\be\label{main}
W = x_1^d + x_2^d \,,
\ee
which will turn out to exhibit interesting and novel phenomena. 
One of the reasons why this case is likely to be significantly
different comes from the fact that it has $c^{\rm total}>3$ (for 
$d\geq 5$). Regarded as a theory with respect to the diagonal $N=2$
algebra (this is the symmetry seen by the matrix factorisation point
of view), the theory is therefore not rational any more. Its brane
spectrum will thus be much richer than for the case of a single minimal
model. 
\smallskip

Another motivation for looking at this example comes from the recent
work of \cite{ADD,ADDF}. They constructed a matrix factorisation both
for a single  D0-brane and a single D2-brane on the quintic. 
Geometrically, the D0 brane is described by the set of equations
\beq\label{introI}
x_1=x_2=x_3=0, \quad x_4-\eta x_5=0 \ ,
\eeq
where $\eta$ is a fifth root of unity. On the other hand, the D2 brane
corresponds to 
\beq\label{introII}
x_1=0, \quad x_2-\eta_1 x_3=0, \quad x_4-\eta_2 x_5=0 \ .
\eeq
As was also shown in \cite{ADD} the corresponding factorisations do 
not correspond to any of the Recknagel-Schomerus (RS) boundary states
\cite{RS}; indeed, it has been  known for some time, that the charges
of the RS boundary states only describe a sublattice of finite index
in the complete charge lattice of the quintic theory \cite{BDLR}.  

Given our analysis of the correspondence for superpotentials of the
form (\ref{main}), we are able to identify the factorisations
corresponding to (\ref{introI}) and (\ref{introII}) with specific
permutation D-branes \cite{R}. In particular, we find stable D0 branes
at the Gepner point for a number of Calabi-Yau manifolds.\footnote{For
the case of the quintic, it was already realised in \cite{R1} that the
relevant permutation boundary state only carries D0-brane charge.}
Furthermore we have checked that these boundary states have all the
required properties: in particular, they carry the correct charges in
the large volume basis (as can be confirmed by relating their charges
to those of the RS branes), and for the case of the D0-brane they
posses three complex marginal operators, corresponding to the motion
of a single D0-brane on the Calabi-Yau manifold.

We have also noted that the intersection matrix of the D0-brane and
its images under the Gepner monodromy always agrees with `the
intersection matrix in the Gepner basis' that had been computed for 
one- and two-parameter examples in \cite{BDLR,DR,KLLW,SchI}. This basis
was obtained by analytic continuation of the fundamental period at
large volume \cite{CDGP,F,KT,CDFKM,CFKM}. In some examples, this basis
already generates the full charge lattice; in other cases (including
for example the quintic), a basis for the complete charge lattice is
generated by the D2-brane and its images under the Gepner monodromy. 
\smallskip

In general the equations (\ref{introI}) do not necessarily describe a
point on the Calabi-Yau. In particular, if the point described by 
(\ref{introI}) lies on a singular locus, it is replaced by an
exceptional set, and the corresponding brane is higher dimensional, a
D2 or D4 brane depending on the geometry.  In such situations one
should expect that the relevant permutation brane carries the
corresponding charge, as we verify in an explicit example.

In most of these examples, the RS branes describe a sublattice of
maximal rank in the full charge lattice. In general, however, the RS
branes do not account for all the charges; in particular, the RS
branes sometimes do not carry the charges of branes that are wrapped
around these exceptional divisors. As an example we consider a 
four parameter-model with two non-toric deformations, for which the RS
branes only span a lower-dimensional sublattice of the charge lattice.
In this model the form of the factorisation suggests that the
permutation branes should carry the missing charges. This can be
verified by showing that the charge lattice generated by the
permutation branes has indeed full rank. 
On the other hand, the permutation branes 
(and the tensor product branes)
do not always account for the full charge lattice as we also
demonstrate with an example. In this case the failure of the
permutation branes to generate the full charge lattice has a simple
geometrical interpretation.    
\medskip

The paper is organised as follows. In section 2 we briefly review the
analysis for the case of a single minimal model. Section~3 describes 
some of the matrix factorisations for the case of the tensor product
of two minimal models, and studies their properties. For a restricted
class of factorisations we propose corresponding boundary states in
section~4, where a number of consistency checks are spelled out. The
application  to the construction of the D0-brane and D2 brane in 
Gepner models is given in section~5, and their geometric
interpretation is discussed in section~6. There are various appendices 
where some of the more technical material is given.
\bigskip

{\it Note added:} While we were in the process of writing up this paper,
we were made aware of related work \cite{ERR}. 
After completion of
this paper, the paper \cite{EGJ} appeared in which the relation
between certain matrix factorisations and geometry is analysed using
orbifold techniques.

\section{The baby example: a single minimal model}

Let us begin by briefly reviewing the correspondence for the case of a
single minimal model \cite{BHLS,KL2,KL3} which corresponds to the
superpotential 
\be
W = x^d \,.
\ee
The corresponding conformal field theory is described by a single
$N=2$ minimal model with $d=k+2$. (Our conventions for the $N=2$
minimal models are summarised in appendix~A.) The spectrum of this
theory is (after GSO-projection)
\be\label{single}
\H = \bigoplus_{[l,m,s]} \,
\left( \H_{[l,m,s]} \otimes \bar\H_{[l,m,-s]} \right) \,.
\ee
This GSO-projection is the analogue of the Type 0A projection; there
is also another GSO-projection (the analogue of Type 0B) for which 
the right-movers lie in the representation $[l,m,s]$ rather than
$[l,m,-s]$. Since we think of embedding this model into a critical
string theory we can always take either of the two GSO-projections for
this internal theory, as long as we compensate this by taking the
appropriate GSO-projection for the remaining degrees of freedom. As we
shall see, the D-branes we are about to construct lie in
(\ref{single}). 

\noindent We are interested in B-type gluing conditions
\ba
\left(L_n - \bar{L}_{-n} \right) |\!| B \rangle\!\rangle & = & 0
\nonumber \\
\left(J_n + \bar{J}_{-n} \right) |\!| B \rangle\!\rangle & = & 0
\label{gluing} \\
\left(G^\pm_r + i\, \eta\, \bar{G}^\pm_{-r} \right) 
|\!| B \rangle\!\rangle & = & 0 \,. \nonumber
\ea
Here $\eta=\pm 1$ describes the two spin-structures. The corresponding
Ishibashi states are then supported in the sectors 
$[l,m,s]\otimes [l,-m,-s]$. For the above spectrum (\ref{single})  we
therefore have Ishibashi states $|[l,0,s]\rangle\!\rangle$ in each
sector $[l,0,s]$ with $l+s$ even; in total there are therefore
$2(k+1)$ Ishibashi states. [This discussion is appropriate for the
bosonic subalgebra of the $N=2$ algebra; if we think in terms of the
$N=2$ symmetry, then the two representations $[l,m,s]$ and $[l,m,s+2]$
form one $N=2$ representation. There are therefore only $k+1$
different $N=2$ representations, but we can choose the two different
spin structures $\eta=\pm 1$ in each case, and therefore there are
also $2(k+1)$ $N=2$ Ishibashi states. Half of them have $\eta=+1$, the
other half $\eta=-1$.]  

The corresponding B-type boundary states were constructed some time
ago (see for example \cite{MMS}), and are explicitly given as  
\be
|\!| L,S \rangle\!\rangle = 
\sqrt{k+2} \,
\sum_{l+s\in 2\Zop} 
\frac{S_{L0S,l0s}}{\sqrt{S_{l0s,000}}} \, 
| [l,0,s]\rangle\rangle \,.
\ee
Here $L=0,1,\ldots,k$ and $S=0,1,2,3$. The boundary states with $S$
even (odd) satisfy the gluing conditions with $\eta=+1$ ($\eta=-1$); 
in the following we shall restrict ourselves to the case $\eta=+1$,
and thus to even $S$.\footnote{The D-branes corresponding to $\eta=-1$
or $S$ odd preserve a different supercharge at the boundary. The
branes that are described by the different matrix factorisations
however always preserve the {\it same} supercharge at the boundary.} 
We also note that 
\be
|\!| L,S \rangle \!\rangle = |\!| k-L,S+2\rangle\!\rangle
\ee
and thus there are only $k+1$ different boundary states with $\eta=+1$
(and $k+1$ different boundary states with $\eta=-1$). These boundary
states therefore account for all the $N=2$ Ishibashi states. Finally
we note that $|\!| L,S\rangle\!\rangle$ and  
$|\!| L,S+2\rangle\!\rangle$ are anti-branes of one another (since
they differ by a sign in the coupling to the RR sector states). 

\noindent The corresponding open string spectrum can be determined
from the overlap
\ba\label{tenspec}
\langle\!\langle L S & |\!| & q^{\frac{1}{2}(L_0 + \bar{L}_0) 
- \frac{c}{24}} \,
|\!| \, \hat{L} \hat{S} \rangle \!\rangle  \\
&& = \sum_{[l,m,s]}
\Bigl( \delta^{(4)}(\hat{S}-S+s) \, N_{\hat{L}\ l}{}^{L} 
+ \delta^{(4)}(\hat{S}-S+2+s) \, N_{\hat{L}\ k-l}{}^{L} \Bigr)
\chi_{[l,m,s]}(\tilde{q}) \,, \nonumber 
\ea
where $N_{L\ l}{}^{M}$ denotes the level $k$ fusion rules of $su(2)$, 
and $\chi_{[l,m,s]}$ is the character of the coset representation. In
particular, we can read off from this expression how many topological
chiral primary states propagate between two such branes; for example,
between the two branes $|\!|L,0\rangle\!\rangle$ and
$|\!|\hat{L},0\rangle\!\rangle$ we have as many chiral primary states
$[l,l,0]$ as there are $l$ for which $N_{\hat{L} \ l}^{L} = 1$.  

These results can now be compared with the analysis based on matrix
factorisations \cite{BHLS,KL3}. The corresponding factorisations of
$W=x^d$ are  
\be
Q_r = 
\begin{pmatrix}
0 & x^r \cr x^{d-r} & 0
\end{pmatrix}\,,
\qquad
J = x^r\,, \qquad E=x^{d-r} \,,
\ee
where $r=1,2,\ldots,d-1$. (As was shown in \cite{HLL1}, all
factorisations of $W=x^d$ are equivalent to direct sums of these
one-dimensional factorisations.) The dictionary between the two
approaches is then    
\be
Q_r \Longleftrightarrow |\!| r-1,0\rangle\!\rangle  \,.
\ee
In particular, the two factorisations $Q_r$ and $Q_{d-r}$ that are
related by interchanging the roles of $E$ and $J$ correspond to
anti-branes of one another.

This relationship can be confirmed by comparing the topological open
string spectra between two such branes. From the point of view of
matrix factorisations this amounts to finding $\phi_0$ and $\phi_1$
such 
that the diagram 
\begin{center}
\begin{picture}(400,110)(-100,20)
\put(00,100){$Q$}
\put(40,100){$\BC[x]$}
\put(80,108){\vector(1,0){50}}
\put(105,113){\footnotesize{$J$}}
\put(150,100){$\BC[x]$}
\put(130,100){\vector(-1,0){50}}
\put(105,90){\footnotesize{$E$}}
\put(00,30){$\hat{Q}$}
\put(40,30){$\BC[x]$}
\put(80,38){\vector(1,0){50}}
\put(105,43){\footnotesize{$\hat{J}$}}
\put(150,30){$\BC[x]$}
\put(130,30){\vector(-1,0){50}}
\put(105,20){\footnotesize{$\hat{E}$}}
\put(50,90){\vector(0,-1){45}}
\put(160,90){\vector(0,-1){45}}
\put(30,65){$\phi_0$}
\put(170,65){$\phi_1$}
\put(70,90){\vector(2,-1){75}}
\put(140,90){\vector(-2,-1){75}}
\put(75,75){$t_0$}
\put(125,75){$t_1$}
\end{picture}
\end{center} 
commutes. Here $\phi_0$ and $\phi_1$ are again polynomials in $x$, and
the commutativity simply means that 
\begin{eqnarray}
0 & = & (D\phi)_0 =\hat{J}\, \phi_0 - \phi_1\, J \,, \nonumber \\
0 & = & (D\phi)_1 =\hat{E}\, \phi_1 - \phi_0\, E \,.
\end{eqnarray}
More abstractly, this is the condition that the morphism defined by
\be
\Phi = \left( \begin{matrix}
\phi_0 & 0 \\ 0 & \phi_1 
\end{matrix} \right)
\ee
is $Q$-closed. In addition, $\Phi$ has to respect the U(1) grading
(see \cite{HW,ADD} for a detailed discussion of this point); this is
trivial for the case in question, but in general imposes a non-trivial
constraint. 

The actual topological states are
described by the $Q$-cohomology \cite{KL1,BHLS,KL2,KL3,HW}; thus we need 
to determine the solutions $(\phi_0,\phi_1)$ up to $Q$-exact
solutions. In the current context the $Q$-exact solutions are  
\begin{eqnarray}
\tilde{\phi}_0 & = & (Dt)_0 = 
 \hat{E}\, t_0 + t_1 \, J  \,, \nonumber \\
\tilde{\phi}_1 & = & (Dt)_1 = \hat{J} \, t_1 + t_0 \, E  \,.
\end{eqnarray}
The $Q$-cohomology of $(\phi_0,\phi_1)$ describes then the `bosonic'
open string degrees of freedom, {\it i.e.} the topological open string
states between the branes corresponding to $Q$ and $\hat{Q}$. The
`fermionic' degrees of freedom, {\it i.e.} the topological open string
states between the brane $Q$ and the anti-brane of $\hat{Q}$, can be
deduced from this by exchanging the roles of $\hat{E}$ and
$\hat{J}$. These degrees of freedom are then described by
$(t_0,t_1)$. Their $Q$-closedness condition is 
\begin{eqnarray}
0 & = & (Dt)_0 = \hat{E}\, t_0 + t_1 \, J  \nonumber \\
0 & = & (Dt)_1 = \hat{J} \, t_1 + t_0 \, E \,, 
\end{eqnarray}
and the $Q$-exact states are those that are of the form 
\begin{eqnarray}
\tilde{t}_0 & = & 
(D\phi)_0 = \hat{J}\, \phi_0 - \phi_1\, J \,, \nonumber \\
\tilde{t}_1 & = & 
(D\phi)_1 = \hat{E}\, \phi_1 - \phi_0\, E  \,.
\end{eqnarray}

For example, for $Q=\hat{Q}=Q_r$, the $Q$-closed condition for the
bosonic degrees of freedom is simply $\phi_0=\phi_1$. The $Q$-exact
solutions are those for which $\phi_0=\phi_1$ contains $x^r$ (or
$x^{d-r}$) as a factor. Thus for $r\leq (d-1)/2$  we have
$r$ different topological states, corresponding to  
$\phi_0=\phi_1=1,x,\ldots, x^{r-1}$. This agrees then precisely with
the topological open string spectrum of $|\!|r-1,0\rangle\!\rangle$
where the chiral primaries $[l,l,0]$ with $l=0,2,\ldots, 2(r-1)$
appear. The analysis for other combinations of branes works
likewise. One can also check that the U(1)-charges match.

\section{The product theory}

Now we want to consider the product theory that corresponds to the
superpotential 
\be
W = x_1^d + x_2^d \,.
\ee
The space of states of the corresponding conformal field theory is
(after GSO-projection)
\ba
\H & = & \bigoplus_{[l_1,m_1,s_1],[l_2,m_2,s_2]}
\Bigl( \left(\H_{[l_1,m_1,s_1]} \otimes \H_{[l_2,m_2,s_2]} \right) 
\otimes \left( \bar\H_{[l_1,m_1,s_1]} \otimes \bar\H_{[l_2,m_2,s_2]}
\right) \label{GSO1} \\
& & \qquad  \qquad
\oplus \,\, 
\left(\H_{[l_1,m_1,s_1]} \otimes \H_{[l_2,m_2,s_2]} \right) 
\otimes \left( \bar\H_{[l_1,m_1,s_1+2]} \otimes 
\bar\H_{[l_2,m_2,s_2+2]} \right) \Bigr) \,, \nonumber
\ea
where the sums over $s_1$ and $s_2$ are restricted to 
$s_1-s_2\in 2\Zop$. [Again, there is also another GSO-projection, but
as we shall see the branes we are about to construct will lie in
(\ref{GSO1}).]   

This theory has the obvious tensor product branes that satisfy the
gluing conditions (\ref{gluing}) separately for the two $N=2$
theories; they are explicitly given by\footnote{We are ignoring in the
following the resolved branes that appear for $L_1=L_2=k/2$ if $k$ is
even.} 
\ba\label{tbr}
|\!|L_1,S_1,L_2,S_2\rangle\!\rangle & = & 
\frac{(2k+4)}{4\sqrt{2}} \, 
\sum_{s_1,s_2}\, \sum_{l_1,l_2} 
\left(
\frac{S_{L_1 0S_1, l_1 0 s_1}}{\sqrt{S_{000,l_1 0 s_1}}}
+ \frac{S_{k-L_1 0 S_1+2, l_1 0 s_1}}{\sqrt{S_{000,l_1 0 s_1}}}
\right) 
\nonumber \\
& & \qquad \times
\left(\frac{ S_{L_2 0 S_2,l_2 0 s_2}}{\sqrt{S_{000,l_2 0 s_2}}} 
+  \frac{ S_{k-L_2 0 S_2+2,l_2 0 s_2}}{\sqrt{S_{000,l_2 0 s_2}}}  
\right)\, 
|[l_1,0,s_1]\otimes [l_2,0,s_2]\rangle\!\rangle \,, \nonumber 
\ea
where $|[l_1,0,s_1]\otimes [l_2,0,s_2]\rangle\!\rangle$ is now the
Ishibashi state in the sector 
\be
|[l_1,0,s_1]\otimes [l_2,0,s_2]\rangle\!\rangle \in 
\Bigl(\H_{[l_1,0,s_1]}\otimes  \H_{[l_2,0,s_2]} \Bigr) \otimes 
{\Bigl(\bar\H_{[l_1,0,-s_1]} \otimes \bar\H_{[l_2,0,-s_2]}\Bigr)}\,,
\ee
and the sum over $l_i$ and $s_i$ is unrestricted, except
that $s_1-s_2$ is even. As before $S_i$ even (odd) describes the
boundary states that satisfy the gluing condition for the $i$th $N=2$
algebra with $\eta=+1$ ($\eta=-1$); we shall therefore
restrict ourselves to considering $S_1$ and $S_2$ even. Furthermore,
the labels $(L_i,S_i)$ are only  defined up to the equivalence
$(L_i,S_i) \sim (k-L_i,S_i+2)$ and $(S_1,S_2)\sim (S_1+2,S_2+2)$. The
open string spectrum between two such branes is essentially given by
the tensor product of  
(\ref{tenspec}) 
\ba
& \langle\!\langle & L_1,S_1,L_2,S_2 |\!| 
q^{\frac{1}{2}(L_0+\bar{L}_0) - \frac{c}{12}} 
|\!|  \hat{L}_1,\hat{S}_1,\hat{L}_2,\hat{S}_2]\rangle\!\rangle  
= \sum_{r=0}^{1}\, \sum_{[l_1 m_1 s_1],[l_2 m_2 s_2]} \nonumber \\
&&  \quad \left( 
\delta^{(4)}(\hat{S}_1 - S_1 + s_1+2r) \, N_{\hat{L}_1 l_1}{}^{L_1} 
+ \delta^{(4)}(\hat{S}_1 +2 - S_1 + s_1+2r) \, N_{k- \hat{L}_1 l_1}{}^{L_1} 
\right) \times  \nonumber \\
& & \quad 
\left( 
\delta^{(4)}(\hat{S}_2 - S_2 + s_2-2r) \, N_{\hat{L}_2 l_2}{}^{L_2} 
+ \delta^{(4)}(\hat{S}_2 +2 - S_2 + s_2-2r) \, N_{k- \hat{L}_2 l_2}{}^{L_2} 
\right) \, \times \nonumber \\
& & \qquad \qquad \quad \times \chi_{[l_1,m_1,s_1]}(\tilde{q})\, 
\chi_{[l_2,m_2,s_2]}(\tilde{q}) \,. \nonumber 
\ea
As before, $(L_1,S_1,L_2,S_2)$ and $(L_1,S_1+2,L_2,S_2)$ are 
anti-branes of one another.  For future reference we also note that
these tensor product branes do {\it not} couple to any RR ground
states; they therefore do not carry any RR charge.      

\noindent These boundary states correspond to tensor products of the
one-dimensional factorisations we considered before \cite{ADD,HW}. More
specifically, the boundary state $|\!|L_1,0,L_2,0\rangle\!\rangle$
corresponds to the factorisation
\be\label{tensorfac}
\xymatrix{
\BC[x_1,x_2]^{\oplus 2} \ar@<0.6ex>[r]^{J} &\BC[x_1,x_2]^{\oplus 2} 
\ar@<0.6ex>[l]^{E}
}\,, 
\qquad \hbox{with} \qquad
J = \begin{pmatrix}
J_2 & J_1 \cr E_1 & - E_2
\end{pmatrix}\,, \qquad
E = \begin{pmatrix}
E_2 & J_1 \cr E_1 & - J_2 
\end{pmatrix} 
\ee
and $J_i = x_i^{L_i+1}$ and $E_i=x_i^{d-1-L_i}$,
such that $W=E_1J_1+E_2J_2$. This follows
essentially from the same analysis as for the case of a single minimal
model. 
\medskip

In addition to these tensor product factorisations, there are however
also rank $1$ factorisations \cite{ADD}. The corresponding boundary
conditions had been studied previously using Landau-Ginzburg techniques
in \cite{GJS}. These factorisations make use of the fact that the
superpotential can be factorised as  
\be
W=\prod_{\eta} (x_1-\eta x_2)\,,
\ee
where $\eta$ is in turn each of the $d$ different $d$'th roots of
$-1$. We can label these roots as 
\be
\eta_m =e^{-\pi i\frac{2m+1}{d}}\,,
\ee
where $m=0,1,\ldots, d-1$. Let us define $D=\{ 0,...,d-1 \}$. Then we
can construct the rank $1$ factorisations
\be
\xymatrix{
\BC[x_1,x_2] \ar@<0.6ex>[r]^{J} &\BC[x_1,x_2] 
\ar@<0.6ex>[l]^{E}
}\,,
\ee
where 
\be
J = \prod_{m \in I} (x_1 - \eta_m x_2)\,, 
\qquad 
E= \prod_{n\in D \setminus I} (x_1-\eta_n x_2) \,.
\ee
Here $I$ is any subset of $D$. 

At this stage it is not clear what $N=2$ boundary states these
factorisations correspond to (we shall make a proposal for at least
some of these factorisations in section~4). In order to be able to
make an identification, we should obtain as much information about
them as possible. In particular we need to determine the open string
spectra involving these rank $1$ factorisations. As we shall see, the
result will have a simple geometric interpretation, so it may be
helpful to review first the relation between matrix factorisations and
geometry.

\subsection{Matrix factorisations and geometry}

In \cite{O} Orlov showed that the category of D-branes 
in Landau Ginzburg theories is equivalent to a certain  
geometrical category $D_{Sg}(X)$ that is non-trivial only on singular 
varieties $X$. To be more precise, topological B-type D-branes on a
variety $X$ correspond geometrically to the bounded derived category
of coherent sheaves on $X$. On a smooth variety, any such sheaf has a
finite  locally free resolution. This is no longer the case on a
singular variety, which was the motivation in \cite{O} to introduce
$D_{Sg}(X)$ as the quotient of the bounded derived category of
coherent sheaves by the subcategory of finite complexes of locally
free sheaves. To understand the relationship with Landau-Ginzburg
models, in particular in our example, we  start from a Landau 
Ginzburg potential $W: \BC^{n} \to \BC$ with an isolated critical
point at the origin, in our case, $W=x_1^d+x_2^d:\BC^2\to \BC$.
Denoting the fiber of $W$ over $0$ by $S_0$, \cite{O} allows us to 
establish a relation between $D_{Sg}(S_0)$ and the Landau-Ginzburg
category. For this, we associate with any factorisation 
$$
\Bigl(
\xymatrix{
P_1 \ar@<0.6ex>[r]^{J} &P_0 \ar@<0.6ex>[l]^{E}}
\Bigl)
$$
the short exact sequence
\begin{equation}\label{shseq}
0\lto P_1 \stackrel{J}{\lto} P_0 \lto \Coker J \lto 0\,.
\end{equation}
The geometrical object associated to the factorisation is then the 
sheaf $\Coker J$, which, since it is annihilated by $W$, is 
a sheaf on $S_0$. Let us apply this to the simplest rank $1$
factorisation, where $J$ is a single linear factor 
$J=x_1-\eta x_2$. $\Coker J$ is then simply the ring 
$\BC[x_1,x_2] / J$, or, geometrically, the line with the
equation  
$$
x_1-\eta x_2 =0\,.
$$
For higher order $J$, $J = \prod_{m \in I} (x_1-\eta_m x_2)$, we  
obtain accordingly a union of lines 
$$
\bigcup_{m\in I} \ \ \{ x_1- \eta_m x_2 =0 \} \,.
$$

Based on this geometric picture, we can make a prediction for the
number of fermionic (and bosonic) operators between pairs of branes. The
idea is simply that the number of fermions corresponds to the number
of intersections between the two branes. For example, on a single
brane, there should be no fermions, whereas between two branes
corresponding to different single lines ($J$ linear) there should be
exactly one fermion as calculated in \cite{ADD}. According to this
logic, between  disjoint sets of lines there should be $d_1 d_2$
fermions, where $d_1$ is the number of lines in the first set, and
$d_2$ the number of lines in the second set. 

We can also count the bosons, since by definition the number of bosons
propagating between two branes is equal to the number of fermions
propagating between the brane and the anti-brane. Exchanging brane and
anti-brane corresponds to exchanging $J$ and $E$ in Landau-Ginzburg
language, which means, in this geometrical language, that the
anti-brane of a brane localised at a given set of lines consists of
the complementary set of lines. Therefore, the number of bosons is
again given by an intersection number, this time between the branes
and anti-branes.  

Finally, the tensor product branes can be thought of as corresponding
to points at the origin, where the number of points is determined by 
the factorisation labels $L_1, L_2$. In particular, they are  
invariant under rotations. This corresponds to the fact that these
branes do not carry RR charge. 
\medskip

We will now compute the open string spectrum from the matrix
factorisation point of view and check that the dimensions 
of the cohomologies indeed match these geometric expectations.

\subsection{Calculating the open string spectrum}

First we consider the open string spectrum between two rank $1$
branes. As a warm-up we consider the case of open strings between a
brane and itself. For the fermions the BRST-invariance condition is 
\be
E\, t_0 + t_1\, J  =0 \,.
\ee
Since $E$ and $J$ do not have any common divisors, the only solution
is $t_1 = a E$ and $t_0= - a J$. It is then easy to see that this
solution is BRST trivial, and thus there are no fermions on a single
rank $1$ brane. 

\noindent For the bosons, the BRST-invariance condition is 
\be
J \, \phi_0 = \phi_1 \, J \,, \qquad E\, \phi_1 = \phi_0 \, E \,,
\ee
from which it can be concluded that $\phi_0=\phi_1$. The boson is BRST
trivial if 
\be
\phi_0 = E\, t_0 + t_1 \, J  \,, \qquad
\phi_1 = J\, t_1 + t_0 \, E \,, 
\ee
which means that the bosonic spectrum is given by the ring 
$\BC[x_1, x_2]/I$, where $I$ is the ideal generated by $J$ and
$E$. To calculate the dimension of this space, we note that the
dimension corresponds to the number of intersections of the two curves 
$E=0$ and $J=0$. According to Bezouts theorem, two plane curves
of degrees $d_0$ and $d_1$ will intersect in $d_1d_0$ points, counting 
intersections at infinity and multiplicities. Since in our case there
are no intersections at infinity, the dimension of the ideal is given
by $d_0d_1$. This is therefore in perfect agreement with the geometric
picture outlined above.

The general case can be shown along similar lines; the details of this
calculation are described in appendix~B. There we also give the
calculation for the open string spectrum between a tensor product and
a rank $1$ brane (for the case that the latter has $\hat{J}$ linear).

\subsection{Matrix flows}

The above picture also suggests that the branes corresponding to
higher order $J$ can be obtained as bound states of the branes with
linear $J$. We want to explain now that this is actually correct. 

As was explained in \cite{HLL1} there is a natural notion of
isomorphism of matrix factorisations: two matrix factorisations $Q_1$
and $Q_2$ are isomorphic if $Q_2 = U Q_1 U^{-1}$, where both $U$ and
its inverse $U^{-1}$ are block-diagonal matrices 
\be
U = \begin{pmatrix}
U_1 & 0 \cr 0 & U_2
\end{pmatrix}
\ee
with polynomial entries. In
particular, this condition implies that the spectrum of $Q_1$ and
$Q_2$ relative to any other brane is the same; thus it is indeed
natural to identify such factorisations. This concept is crucial to
understand the bound state formation of branes from the topological
point of view \cite{HLL2}. 

\noindent Suppose now that we are given a pair of branes 
\be
\Bigl(
\xymatrix{
P_0 \ar@<0.6ex>[r]^{J_P} &P_1 \ar@<0.6ex>[l]^{E_P}
}
\Bigl) \quad {\rm and } \quad
\Bigl(
\xymatrix{
O_0 \ar@<0.6ex>[r]^{J_O} &O_1 \ar@<0.6ex>[l]^{E_O}
}
\Bigl) \,,
\ee
whose relative open string contains a tachyon. By this we mean a 
boundary changing operator $t=(t_0,t_1)$, $t_0:P_0 \to O_1$,
$t_1:P_1\to O_0$ that is BRST closed but not BRST exact. 
Likewise, there can be tachyons in the other direction 
$t'=(t'_0,t'_1)$, $t'_0:O_0 \to P_1$ and $t'_1:O_1 \to P_0$.
A bound state of these two branes with that tachyon profile should
have the form
\be\label{bound}
\Bigl(
\xymatrix{
P_0 \oplus O_0\ar@<0.6ex>[r]^{J} &P_1\oplus O_1 \ar@<0.6ex>[l]^{E}
}
\Bigl),
\ee
where the maps $J$ and $E$ are given by
\be
J= \left( \begin{array}{cc} J_P & t_0' \\ t_0 & J_O \end{array} \right),
\quad
E= \left( \begin{array}{cc} E_P & t_1' \\ t_1 & E_O \end{array} \right) \ .
\ee
The BRST operator of the combined system is then
\be
Q=\left( \begin{array}{cc} 0 & J \\ E & 0 \end{array} \right) \,.
\ee
The condition that a bound state is formed in this way is that
(\ref{bound}) is a valid boundary condition fulfilling $Q^2=W$. Using
the above notion of isomorphism, the resulting boundary condition may
in fact be equivalent to one of the other boundary conditions. This
occurs in particular for the case where we have two rank $1$ branes
with complementary factors.

\subsubsection{Rank $1$ flows}

Let us consider the superposition of two rank $1$ branes
\be
J_P=\prod_{n\in I_P} (x_1-\eta_n x_2)\,, \qquad 
E_P = \prod_{n'\in D\setminus I_P}
(x_1-\eta_{n'} x_2) \,,
\ee
and
\be
J_O=\prod_{m\in I_O} (x_1-\eta_{m} x_2)\,, \qquad
E_O = \prod_{m'\in D\setminus I_O}
(x_1-\eta_{m'} x_2)\,,
\ee
where $I_P$ and $I_O$ are disjoint, $I_P\cap I_O = \emptyset$. Then
$E_P$ contains $J_O$ as a factor, and $E_O$ contains $J_P$ as a
factor. As we have explained before, the open string spectrum of this
configuration contains $|I_P|\, |I_O|$ fermions. We can therefore
modify the BRST operator of this configuration by  
\be\label{Qeinf}
Q =  \left(
\begin{matrix}
0 & 0 & J_P & 0 \cr 
0 & 0 & \lambda \,\, & J_O \cr 
E_P & 0 & 0 & 0\cr
- \lambda \tilde{E}\,\, & E_O \,\, & 0 & 0 
\end{matrix}\right) \qquad \tilde{E} = \frac{E_P}{J_O} 
= \frac{E_O}{J_P} 
\,,
\ee
where $\lambda$ labels the different fermionic perturbations. One
easily checks that this $Q$ still squares to the superpotential. 
For constant $\lambda$ one then finds that this BRST operator is
equivalent to the BRST operator 
\be
\hat{Q} = \left(
\begin{matrix}
0 & 0 & 0 & J_P\,J_O \cr
0 & 0 & 1 \,\, & 0 \cr
0 & W \,\, & 0 & 0 \cr
\tilde{E} \,\,& 0 & 0 & 0
\end{matrix} \right) \,.
\ee
In fact, the relevant invertible matrix $U$ that satisfies
$Q\, U = U \, \hat{Q}$ is given by 
\be
U = \left(
\begin{matrix}
a\,\, & J_P\, d \,\,& 0 & 0 \cr
0 & \lambda \, d & 0 & 0 \cr
0 & 0 & d\,\, & a\, J_O \cr
0 & 0 & 0 & - \lambda \, a 
\end{matrix}
\right)\,.
\ee
It is clear that the inverse of this matrix (for constant
$a,d,\lambda$) is again a matrix with polynomial entries. On the other
hand, the BRST operator $\hat{Q}$ simply describes the 
superposition of a trivial brane (corresponding to the trivial matrix
factorisation) with the rank $1$ brane described by 
\be
\tilde{J} = J_P\, J_O =
\prod_{n\in I_P\cup I_O} (x_1-\eta_n x_2)\,, \qquad 
\tilde{E} = \prod_{n'\in D\setminus \{I_P\cup I_O\}}
(x_1-\eta_{n'} x_2) \,.
\ee
This argument therefore shows that two complementary rank $1$ branes
can flow to the rank $1$ brane that is described by their product.  
This thus confirms the geometric picture that was put forward in
section~3.1.

\subsubsection{Flows from rank $1$ to tensor product branes}

The other flow that is of interest relates rank $1$ branes to tensor
product branes. The simplest example concerns the configuration of a
rank $1$ brane (corresponding to a single factor) with its anti-brane,
{\it i.e.} 
\be
J_P= (x_1-\eta_n x_2)\,, \qquad 
E_P = \prod_{n'\ne n}
(x_1-\eta_{n'} x_2) \,,
\ee
and
\be
J_O=\prod_{m\ne n } (x_1-\eta_{m} x_2)\,, \qquad
E_O = (x_1-\eta_{n} x_2)\,.
\ee
This is a special case of the situation considered in the previous
subsection, and the BRST operator is therefore again of the form
(\ref{Qeinf}) with $\tilde{E}=1$. If we take $\lambda$ to be constant,
then the above analysis implies that the configuration flows to the
trivial brane configuration (since $J_P J_O = W$). In order to flow to
a non-trivial brane configuration, we therefore consider 
\be\label{lambdadef}
\lambda = x_1 + \eta_n  \, x_2 \,.
\ee
Then one finds that this BRST matrix is equivalent to 
\be\label{Qhat}
\hat{Q} = 
\left(
\begin{matrix}
{\bf 0} \,\,& r_1 \cr
r_0 & {\bf 0}
\end{matrix}
\right)\,, \qquad
r_1 = \left(
\begin{matrix}
x_2 \,\,& x_1^{d-1} \cr
x_1 & - x_2^{d-1}
\end{matrix}
\right) \,, \qquad
r_0 = \left(
\begin{matrix}
x_2^{d-1} \,\,& x_1^{d-1} \cr
x_1 & - x_2
\end{matrix}
\right) \,, \qquad
\ee
where the relevant $U$-matrix satisfying $Q \, U = U \, \hat{Q}$ is
simply 
\be
U = \left(
\begin{matrix}
-\eta_n\, d \,\,& d\,\, & 0 & 0 \cr
\eta_n\, d & d & 0 & 0 \cr
0 & 0 & d \,\,& -d\, v \cr
0 & 0 & 0 & 2 \eta_n \, d
\end{matrix}
\right)\,,
\qquad
v = \frac{x_2^{d-1} + \eta_n x_1^{d-1}}{x_1 - \eta_n x_2} \,.
\ee
In particular, one observes that $v$ is a polynomial in $x_1$ and
$x_2$ (since $\eta_n$ is a d$th$ root of $-1$), and therefore $U$, as
well as its inverse, are matrices with polynomial entries.

On the other hand, we recognise the BRST matrix (\ref{Qhat}) to be the 
tensor product brane corresponding to 
\be\label{basicten}
J_1 = x_1^{d-1} \,, \qquad J_2 = x_2 \,.
\ee
Repeating this construction we can therefore obtain any of the tensor
product branes from the rank $1$ branes. To this end we observe,
following \cite{HLL2}, that there is a flow relating the tensor
product branes
\be\label{tenflow}
J_1 = x_1^m\,, \, J_2 = x_2  \quad \oplus
\quad 
\tilde{J}_1 = x_1^m \,,\, \tilde{J}_2 = x_2^l \qquad
\longrightarrow \qquad
\hat{J}_1 = x_1^m \,,\, \hat{J}_2 = x_2^{l+1} \,,
\ee
and similarly 
\be\label{tenflow1}
J_1 = x_1^{d-1}\,, \, J_2 = x_2^n  \;\;\oplus
\;\;
\tilde{J}_1 = x_1^{d-l} \,,\, \tilde{J}_2 = x_2^n \quad
\longrightarrow \quad
\hat{J}_1 = x_1^{d-l-1} \,,\, \hat{J}_2 = x_2^{n} \,.
\ee
Combining these two flows, it is then clear that every tensor product 
brane can be obtained from a suitable combination of the tensor
product brane that is described by (\ref{basicten}). In turn, this
last brane could be obtained from two rank $1$ branes. Combining these
arguments, it therefore follows that {\it every} tensor product brane
can be obtained from a suitable combination of the rank $1$ branes.

\section{Permutation branes}

In the previous section we have analysed the properties of the rank
$1$ factorisations in detail. Now we want to make a proposal for which
$N=2$ superconformal boundary states these factorisations correspond
to.

The diagonal $N=2$ algebra (which is the symmetry that is relevant
from the matrix factorisation point of view) has central charge $2c$; 
except for the case of $k=1$ that was discussed in \cite{BHLW}, the
diagonal $N=2$ algebra therefore does not define a minimal
model. Regarded as a theory with respect to this algebra, the theory
is therefore (for $k>1$) not rational. It is then difficult to 
find all $N=2$ boundary states of this theory.\footnote{For the case
$k=2$ for which $2c=3$, the techniques of \cite{GK} should allow one
to find a complete description of these boundary states; this will be
described elsewhere.} However, there are always two classes of
`rational' D-branes one can easily construct: the tensor product
branes we have considered at the beginning of section~3, and the
permutation branes \cite{R} (see also \cite{GSS}).

\noindent The permutation branes are characterised by the gluing
conditions 
\ba
\left(L^{(1)}_n - \bar{L}^{(2)}_{-n} \right) 
|\!| B \rangle\!\rangle =
\left(L^{(2)}_n - \bar{L}^{(1)}_{-n} \right) 
|\!| B \rangle\!\rangle 
& = & 0
\nonumber \\
\left(J^{(1)}_n + \bar{J}^{(2)}_{-n} \right) |\!| B \rangle\!\rangle 
= \left(J^{(2)}_n + \bar{J}^{(1)}_{-n} \right) |\!| B \rangle\!\rangle 
& = & 0 
\label{gluingp} \\
\left(G^{\pm (1)}_r + i \eta_1 \bar{G}^{\pm (2)}_{-r} \right) 
|\!| B \rangle\!\rangle = 
\left(G^{\pm (2)}_r + i \eta_2 \bar{G}^{\pm (1)}_{-r} \right) 
|\!| B \rangle\!\rangle
& = & 0 \,. \nonumber
\ea
Provided that $\eta_1=\eta_2$, these gluing conditions imply that the
diagonal $N=2$ gluing conditions are respected. For the theory under
consideration (\ref{GSO1}) we have permutation Ishibashi states in the
sectors  
\be
|[l,m,s_1]\otimes [l,-m,-s_2]\rangle\!\rangle^\sigma  \in   
\Bigl(\H_{[l,m,s_1]}\otimes \H_{[l,-m,-s_2]}\Bigr) \otimes
\Bigl(\bar\H_{[l,m,s_2]}\otimes \bar\H_{[l,-m,-s_1]}\Bigr) \,.
\ee
The corresponding boundary states are then 
\ba
|\!| & [ & L,M,S_1,S_2\ ]\ \rangle\!\rangle \label{permstates}\\
& & = \frac{1}{2\, \sqrt{2}} \sum_{l,m,s_1,s_2} 
\frac{S_{Ll}}{S_{0l}} \, e^{i\pi M m / (k+2)}\, 
e^{-i\pi (S_1 s_1 - S_2 s_2)/2}\,
|[l,m,s_1]\otimes [l,-m,-s_2]\rangle\!\rangle^\sigma  
\,,\nonumber 
\ea
where the sum runs over all $l,m,s_1$ and $s_2$ for which 
\be
l+m+s_1 \quad \hbox{and} \quad s_1-s_2 \quad \hbox{are even.}
\ee
The labels $[L,M,S_1,S_2]$ are defined for $L+M + S_1-S_2$ even
only. Again, $S_1$ and $S_2$ correspond to the choice of the spin
structures $\eta_1$ and $\eta_2$, respectively; in order to preserve
the diagonal $N=2$ algebra we therefore need that $S_1-S_2$ is even,
in which case also $L+M$ is even. As before we shall only consider the
case that  $\eta_1=\eta_2=+1$, {\it i.e.} that both $S_1$ and $S_2$
are even. Note that the boundary state is invariant under replacing
both $S_1$ and $S_2$ by $S_i\mapsto S_i+2$. Furthermore we have the 
equivalence $[L,M,S_1,S_2]\simeq [k-L,M+k+2,S_1+2,S_2]$. The branes 
with $(S_1,S_2)$ and $(S_1+2,S_2)$ are anti-branes of one another.

In contradistinction to the tensor product branes, these permutation
branes now couple to the RR ground states. In fact, the coefficient of
the RR ground states in the sector
\be
\left(\H_{[l,l+1,1]}\otimes \H_{[l,-l-1,-1]} \right) \otimes
\left(\bar\H_{[l,l+1,1]}\otimes \bar\H_{[l,-l-1,-1]} \right)
\ee
in the boundary state labelled by $[L,M,S_1,S_2]$ is precisely
\be
Q_{l}\left(|\!|[L,M,S_1,S_2]\rangle\!\rangle \right) = 
\frac{1}{\sqrt{2}}\, 
e^{i\pi S_2/2}\, \frac{S_{LMS_1,l(l+1)1}}{S_{000,l(l+1)1}} \,.
\ee
For the following it is also useful to understand the behaviour of
these permutation branes under the $\Zop_{k+2}$ axial symmetry.
Recall that each minimal model has a $\Zop_{k+2}$ symmetry, whose
generator $g$ acts on the states in 
$\H_{[l,m,s]}\otimes \bar{\H}_{[l,m,s']}$ as 
\be\label{gacdef}
\left. g\right|_{\H_{[l,m,s]}\otimes \bar\H_{[l,m,s']}}
= \exp\left(2\pi i \frac{m}{k+2} \right) \,.
\ee
(Thus $g$ acts as the simple current $[0,2,0]$.) It is easy to see
from the explicit formula (\ref{permstates}) that 
\be\label{permtra}
g_1\, |\!| [L,M,S_1,S_2]\rangle\!\rangle = 
|\!| [L,M+2,S_1,S_2]\rangle\!\rangle  \,, \quad
g_2\, |\!| [L,M,S_1,S_2]\rangle\!\rangle = 
|\!| [L,M-2,S_1,S_2]\rangle\!\rangle \,.
\ee
In particular, the permutation boundary state is therefore invariant
under $g_1\,g_2$. 

\subsection{The dictionary}

We are now in the position to identify a subset of the rank $1$
factorisations with permutation branes. The precise correspondence is
as follows:
\be\label{main1}
|\!|[L,M,S_1=0,S_2=0]\rangle\!\rangle \qquad \Longleftrightarrow 
\qquad J = \prod_{m=(M-L)/2}^{(M+L)/2} (x_1 - \eta_m x_2) \,.  
\ee
For $|\!|[L,M,S_1=2,S_2=0]\rangle\!\rangle = 
|\!|[L,M,S_1=0,S_2=2]\rangle\!\rangle$ the roles of $J$ and $E$ are
interchanged. We should note that this identifies only a subset of the
rank $1$ factorisations with permutation branes (since the phases that
appear on the right hand side are `consecutive'); it would be very
interesting to understand how to describe the remaining factorisations
in terms of conformal field theory. On the other hand, our proposal
does account for all rank $1$ factorisations with $J$ linear --- these
are precisely the permutation branes with $L=0$. Since these
factorisations generate (upon forming bound states) all the
factorisations we have considered, we are at least accounting for all
the RR charges. Also, as we shall see, these are the factorisations
that are relevant for the construction of the D0 and D2-brane in
Gepner models.

Before we begin checking this proposal in some detail, it is useful to
observe that it transforms at least correctly under the axial
symmetries. The $\Zop_{k+2}$ symmetry $g_i$ in conformal field theory
corresponds, in terms of matrix factorisations, to the maps  
\be
g_i: \qquad x_i \mapsto e^{\frac{2\pi i}{k+2}}\, x_i \,.
\ee
Under $g_i$, each factor $(x_1-\eta_n x_2)$ therefore gets mapped to
\be
g_1 (x_1-\eta_n x_2) = e^{\frac{2\pi i}{k+2}} (x_1 - \eta_{n+1} x_2) 
\ee
and 
\be
g_2 (x_1-\eta_n x_2) = (x_1 - \eta_{n-1} x_2) \,,
\ee
respectively. Since overall factors do not matter, we therefore see
that $g_1$ shifts $M\mapsto M+2$ on the right-hand-side of
(\ref{main1}), while $g_2$ acts as $M\mapsto M-2$. This is then
precisely in accord with the transformation properties of the
permutation branes of (\ref{permtra}). 
\smallskip

There are three additional sets of consistency checks for this
identification that we have performed; they will now be described in
turn.

\subsection{Open string spectrum between permutation branes}

It is straightforward to calculate the corresponding open
string spectrum, and one finds 
\ba
& \langle\!\langle & [L,M,S_1,S_2] |\!|  
q^{\frac{1}{2}(L_0 + \bar{L}_0) - \frac{c}{12}}
\,\, |\!| \,\, [\hat{L},\hat{M},\hat{S}_1,\hat{S}_2] \rangle\!\rangle   
=  \sum_{[l_i',m_i',s_i']}  
\chi_{[l_1',m_1',s_1']}(\tilde{q})\, 
\chi_{[l_2',m_2',s_2']}(\tilde{q}) \nonumber
\\
& & \sum_{\hat{l}}
\Bigl[ N_{\hat{l} \hat{L}}{}^{L} \, N_{l_1' l_2'}{}^{\hat{l}}\,
\delta^{(2k+4)}(\Delta M+m'_1-m'_2) \nonumber \\
& & \qquad \times
\Bigl( \delta^{(4)}(\Delta S_1+s_1')\, \delta^{(4)}(\Delta S_2+s_2')
+ \delta^{(4)}(\Delta S_1 +2+s_1')\, \delta^{(4)}(\Delta S_2 +2+s_2') 
\Bigr) \nonumber \\
& & \quad + N_{\hat{l}\, k-\hat{L}}{}^{L} \, N_{l_1' l_2'}{}^{\hat{l}}\,
\delta^{(2k+4)}(\Delta M+k+2+m'_1-m'_2) \nonumber \\
& & \qquad \times 
\Bigl( \delta^{(4)}(\Delta S_1+2+s_1')\, \delta^{(4)}(\Delta S_2+s_2') 
+ \delta^{(4)}(\Delta S_1+s_1')\, \delta^{(4)}(\Delta S_2+2+s_2') 
\Bigr) \Bigr] \,, \nonumber
\ea
where $\Delta M = \hat{M} - M$ and $\Delta S_i = \hat{S}_i-S_i$. 

Let us first consider the case $\Delta_i S = 0$, {\it i.e.} the
overlap between branes and branes. (In the language of matrix
factorisations these are the `bosons'.) Since the sum runs over
equivalence classes $[l_i',m_i',s_i']$ and only even values of $s_i'$
contribute, we may restrict ourselves, without loss of generality, to
$s_1'=s_2'=0$. The chiral primaries are then characterised by
$l_i'=m_i'$. Since $m_1'-m_2'=M-\hat{M}$, and $l_2'$ is contained in
the fusion product of $L$, $\hat{L}$ and $l_1'$, we therefore get  
one topological chiral primary for each $l_1'$ and $l_2'$ for which  
\be\label{boscond}
{\rm {\bf boson}:}\qquad l_2'=l_1' \pm (M-\hat{M}) \;\; {\rm mod}\;\;
(2k+4) 
\qquad \hbox{and} 
\qquad l_2' \subset L \otimes \hat{L} \otimes l_1' \,.
\ee
In order to determine the topological `fermions', we need to consider
the overlap between a brane and an anti-brane, {\it i.e.} 
$\Delta S_1=2$ and $\Delta S_2=0$, say. Then an identical analysis
leads to 
\be
{\rm {\bf fermion}:} \qquad 
l_2' = - l_1'-2 \pm (M-\hat{M}) \;\; {\rm mod}\;\; (2k+4) 
\qquad \hbox{and} \qquad
l_2' \subset L \otimes \hat{L} \otimes l_1' \,.
\ee
It follows from the first equation that there are no bosons if 
$L+\hat{L}<|M-\hat{M}|$. [Here, as in the following, we shall assume that
the $M_i$ have been chosen such that $|M-\hat{M}|\leq k+2$.]
On the other hand, if $L+\hat{L}=|M-\hat{M}|$, then there are
precisely  $k+1-(L+\hat{L})$ bosons; to see this one observes that
only the representation $L+\hat{L}$ in the fusion product of $L$ and
$\hat{L}$ can contribute in (\ref{boscond}), and that one then gets
one solution each for $l_1'=0,1,2,\ldots,k-(L+\hat{L})$. If 
$L+\hat{L}=|M-\hat{M}|+2$ we have in addition also 
$k+3-(L+\hat{L})$ solutions coming from the $L+\hat{L}-2$ term in the  
fusion product of $L$ and $\hat{L}$, leading to  
$k+1-(L+\hat{L}) + k+3-(L+\hat{L})$ boson states. In the general case
when $L+\hat{L}=|M-\hat{M}|+2U$ we therefore have 
\be\label{ppfin}
\sum_{d=0}^{U} \left(k+1 - (L+\hat{L}) + 2d \right) = 
(U+1) \left( k + 1 + U - (L+\hat{L}) \right)
\ee
topological bosonic states. (Here we have assumed that 
$|M-\hat{M}|>|L-\hat{L}|$ --- the other cases can be treated
similarly.) The number of fermions is given by the 
same formulae, except that we have to replace $L$ by $(k-L)$.

These results now need to be compared with the results of the
calculation in appendix~B, in particular, (\ref{mppfin}) and 
(\ref{fermtop}). Using the above identification (\ref{main1}),
$L$ and $\hat{L}$ are related to $I$ and $\hat{I}$ as 
\be
L = |I|-1\,, \qquad \hat{L} = |\hat{I}|-1 \,.
\ee
Furthermore, $U<0$ corresponds precisely to the 
case $|I\cap \hat{I}| = \emptyset$. If this is the case, the number of
topological boson states vanishes, in agreement with (\ref{mppfin}). 
On the other hand, if $U\geq 0$ we have the relation 
(we are assuming here that $I$ and $\hat{I}$ are not subsets of
each other --- this is the analogue of the condition 
$|M-\hat{M}|>|L-\hat{L}|$)
\be
U = |I \cap \hat{I}| -1 \,.
\ee
Then (\ref{ppfin}) becomes
\be
|{\rm bosons}| = |I\cap \hat{I}| \, |D \setminus \{I\cup \hat{I}\}|
\,, 
\ee
and therefore agrees precisely with (\ref{mppfin}). The number of
topological fermions can be obtained from either description upon
replacing $L$ by $k-L$, and their number therefore also agrees.

\subsection{Open string spectrum between permutation and tensor
branes} 

The calculation of the open string spectrum between a permutation and
a tensor product brane\footnote{This is where our analysis differs
  from \cite{R}.} is actually quite subtle. The subtle point  
concerns the calculation of the overlap between the tensor product
Ishibashi state $|[l,0,s]\otimes [l,0,s]\rangle\!\rangle$
and the permutation Ishibashi state in the same sector,
$|[l,0,s]\otimes [l,0,s]\rangle\!\rangle^\sigma$. On general grounds
one knows that this overlap is equal to 
\be
\langle\!\langle [l,0,s]\otimes [l,0,s] | 
q^{\frac{1}{2}(L_0+ \bar{L}_0)-\frac{c}{12}} 
| [l,0,s]\otimes [l,0,s] \rangle\!\rangle^\sigma =
{\rm Tr}_{[l,0,s]\otimes [l,0,s]} 
\left(q^{L_0-\frac{c}{12}} \,\sigma\right) \,. \label{c1}
\ee
Here the trace is taken in the tensor product of 
$\H_{[l,0,s]}\otimes \H_{[l,0,s]}$, and $\sigma$ is the operator that
acts on states in this tensor product by exchanging the two factors. 
To evaluate this trace we observe that only the diagonal terms
contribute. Thus it is clear that the above overlap is proportional to
$\chi_{[l,0,s]}(q^2)$. Now the subtlety concerns the fact that,
depending on the value of $s$, we are dealing with bosonic or
fermionic states. Since $\sigma$ interchanges the states, it picks up
a minus sign in the fermionic case relative to the bosonic case. Thus
we find that 
\be\label{import}
{\rm Tr}_{[l,0,s]\otimes [l,0,s]} 
\left(q^{L_0-\frac{c}{12}} \,\sigma\right) = 
e^{- i\pi s/2}\, \chi_{[l,0,s]}(q^2) \,.
\ee
With this in mind we then calculate that the overlap between a
permutation and a tensor product brane equals  
\ba
&& \langle\!\langle   L_1,S_1,L_2,S_2 |\!| 
q^{\frac{1}{2}(L_0 + \bar{L}_0) - \frac{c}{12}}
\,\, |\!| \,\, [\hat{L},\hat{M},\hat{S}_1,\hat{S}_2] \rangle\!\rangle   
\label{relov} \\
& & \qquad =  \sum_{[l,m,s]}  
\chi_{[l,m,s]}(\tilde{q}^{1/2}) \,
\sum_{\hat{l}} 
\Bigl( N_{L_1 L_2}{}^{\hat{l}}\, N_{\hat{l} \hat{L}}{}^{l} \,
\delta^{(4)}(s+\hat{S}_1+\hat{S}_2 - (S_1+S_2) + 1 ) \nonumber \\
& & \qquad
\qquad \qquad \qquad \qquad \qquad \quad
+ N_{k- L_1 L_2}{}^{\hat{l}}\, N_{\hat{l} \hat{L}}{}^{l} \,
\delta^{(4)}(s+\hat{S}_1+\hat{S}_2 - (S_1+S_2) - 1 ) \Bigr)\,. \nonumber 
\ea
We observe that the representations that appear in the open string
channel are formally R-sector representations; however, they are to be
interpreted as twisted NS-sector representations, where the twist is
again the exchange of the two $N=2$ factors. [Note that if we had left
out the factor of $e^{-i\pi s/2}$ from (\ref{import}), then the open
string representations would have had $s$ even, and the characters
could not have been interpreted in terms of twisted
NS-representations; for a simple example this is explicitly
demonstrated in appendix~C. The fact that the twisted
NS-representations are formally R-sector representations was also
already realised in \cite{FKS}.] 

The fact that $s$ odd appears in these overlaps is also crucial from
the point of view of obtaining the correct topological spectrum. The
characters that appear in (\ref{relov}) are characters of a 
{\it single} $N=2$ minimal model with central charge $c$, evaluated at
$\tilde{q}^{1/2}$. They should however be thought of as NS-characters
of the diagonal $N=2$ algebra whose central charge is $2c$. Thus we
should decompose them as 
\be
\tilde{q}^{1/2(h-{\frac{c}{24}})} + \cdots = 
\tilde{q}^{h^d - \frac{c}{12}} + \cdots \,,
\ee
where $h^d$ is the conformal dimension with respect to the diagonal
algebra. Thus we find
\be\label{komi}
h^d = \frac{1}{2} \, h + \frac{c}{16} \,.
\ee
On the other hand, the U(1)-charge with respect to the diagonal
algebra is just $q^d=q$. The chiral primaries are the states for which
the conformal dimension $h^d$ is half the U(1)-charge $q^d$,
$h^d=q^d/2$. Thus the chiral primaries appear in the representations
where $q=h+c/8$. One easily checks that this is the case if the
representation $(l,m,s)$ is of the form $(l,1-l,1)$ or 
$(l,l+3,-1)$. 

[Incidentally, for the case of the overlap between boundary states of
opposite spin structure, {\it i.e.} for example, $\hat{S}_i=S_i+1$, 
the open string representations can be thought of as describing
(twisted) R-sector representations. In this case $s$ is still odd, and
the chiral primaries correspond to those representations for
which $[l',m',s']$ is a Ramond ground state; indeed, these are the
only states whose contribution to the open string trace is independent
of $\tilde{q}$!]

Thus we have topological states whenever $m'=1-l'$ and $s'=1$ or 
$m'=l'+3$ and $s'=-1$. It is then clear that we get one topological
boson for each representation $l'$ that appears in the fusion product 
\be\label{rescomp}
l' \subset L_1 \otimes L_2 \otimes \hat{L} \,.
\ee
Since $\hat{L}\mapsto k-\hat{L}$ is a simple current, the result for
the fermions is the same. 

This has to be compared now with the calculation of appendix~B, where
we only considered $\hat{L}=0$, {\it i.e.} a permutation brane with
$\hat{J}=x_1-\eta x_2$. For that case, (\ref{rescomp}) reduces simply
to 
\be
\label{relot}
|{\rm bosons}| = |{\rm fermions}| = \min(L_1+1,L_2+1) \,.
\ee
This then agrees precisely with the results of appendix~B, in
particular (\ref{pertenre}).

\subsection{Flows}

Finally, we can check whether the flows we found in section~3.3. from
the matrix factorisation point of view are compatible with the
RR-charges of the conformal field theory description. Let us first
analyse the case discussed in section~3.3.1. In order to be able to
compare this with the conformal field theory results we need to
consider a configuration where all three rank $1$ factorisations have
an interpretation in terms of permutation branes ({\it i.e.}
correspond to `consecutive' lines). Translated into conformal field
theory language, the rank $1$ matrix flows than predict that there is
a flow  
\be\label{flow1}
|\!|[L_1,M_1,0,0]\rangle\!\rangle \oplus 
|\!|[L_2,M_2,0,0]\rangle\!\rangle
\longrightarrow 
|\!|[L_1+L_2+1,M_1+L_2+1,0,0]\rangle\!\rangle \,,
\ee
where 
$M_2 - M_1 - 2 = L_1+L_2$. However, such a flow can only exist if the
RR charges of both sides agree. The $Q_l$ charge of the left-hand side
is   
\begin{eqnarray}
Q_{l}\left(|\!|[L_1,M_1,0,0]\rangle\!\rangle \right) 
& + & Q_{l}\left(|\!|[L_2,M_2,0,0]\rangle\!\rangle \right) \nonumber \\
& = & 
\frac{1}{\sqrt{2}\, S_{0l}} 
\left( S_{L_1 l} e^{i\pi M_1 (l+1) / (k+2)}
+ S_{L_2 l} e^{i\pi M_2 (l+1) / (k+2)} \right) \,.
\end{eqnarray}
Apart from the normalisation factor of the $S$-matrix of $su(2)_k$,
the bracket is therefore 
\begin{eqnarray}
\sqrt{\frac{k+2}{2}}\, (\,\, \cdot\,\, ) & = & 
e^{i\pi (M_1+L_2+1) (l+1) / (k+2)} 
\Bigl[ \sin\left(\frac{\pi (L_1+1) (l+1)}{(k+2)}\right) 
e^{-i\pi (L_2+1)(l+1)/(k+2)} 
\nonumber \\
& & \qquad \qquad
+  \sin\left(\frac{\pi (L_2+1) (l+1)}{(k+2)}\right) 
e^{i\pi (L_1+1)(l+1)/(k+2)} \Bigr]
\nonumber \\
& = & 
e^{i\pi (M_1+L_2+1) (l+1) / (k+2)} \, 
\sin\left(\frac{\pi (L_1+L_2+2)(l+1)}{(k+2)}\right) \,.
\end{eqnarray}
Putting back the various factors we therefore obtain that 
\begin{eqnarray}
Q_{l}\left(|\!|[L_1,M_1,0,0]\rangle\!\rangle \right) 
& + & Q_{l}\left(|\!|[L_2,M_2,0,0]\rangle\!\rangle \right) \nonumber
\\
& = & 
Q_{l}\left(|\!|[L_1+L_2+1,M_1+L_2+1,0,0]\rangle\!\rangle \right)  \,.
\end{eqnarray}
Thus these matrix flows are indeed compatible with the RR charges of
the conformal field theory description!

\subsubsection{$g$-factors}

We can also determine the $g$-factors of the various D-branes from
their boundary state description. The $g$-factor is simply the
coefficient of the Ishibashi state corresponding to
$(l_i,m_i,s_i)=(0,0,0)$. For example, the $g$-factor of the
permutation brane $|\!|[L,M,0,0]\rangle\!\rangle$ equals 
(in the following we are dropping the superfluous $(0,0)$ labels) 
\be
g (|\!|[L,M]\rangle\!\rangle) = \frac{1}{\sqrt{2}}\, 
\frac{S_{L0}}{S_{00}} = \frac{1}{\sqrt{2}}\, 
\frac{\sin(\pi (L+1)/(k+2))}{\sin(\pi/(k+2))} \,.
\ee
On the other hand, the $g$-factor of the tensor product brane 
$|\!|L_1,0,L_2,0\rangle\!\rangle$ is 
\begin{eqnarray}
g(|\!|L_1,L_2\rangle\!\rangle ) & = & 
\frac{(2k+4)}{\sqrt{2}}\, 
\frac{S_{L_100,000}\, S_{L_200,000}}{S_{000,000}}
= \sqrt{k+2}\, \frac{S_{L_10}\, S_{L_20}}{S_{00}} \nonumber \\
& = & \sqrt{2}\, 
\frac{\sin(\pi (L_1+1)/(k+2))\, \sin(\pi (L_2+1)/(k+2))}{\sin(\pi/(k+2))} \,. 
\end{eqnarray}
In particular, we see from these expressions that the flows
(\ref{flow1}) that relate the permutation branes among each other are
{\it perturbative}, {\it i.e.} that the ratio of the $g$-factors of
the initial and final configuration approaches $1$ in the limit
$k\rightarrow \infty$. Indeed, the limit of the relevant ratio is in
that case 
\begin{eqnarray}
\frac{ g(|\!|[L_1+L_2+1,M_1+L_2+1]\rangle\!\rangle)}
{g (|\!|[L_1,M_1]\rangle\!\rangle)
+ g(|\!|[L_2,M_2]\rangle\!\rangle)}
& = & \frac{\sin((L_1+L_2+2) \pi / (k+2))}{
\sin((L_1+1) \pi / (k+2)) + \sin((L_2+1) \pi / (k+2))} \nonumber \\
& & 
\stackrel{k\rightarrow \infty}{\longrightarrow}
\frac{L_1+L_2+2}{L_1+1+L_2+1} = 1 \,.
\end{eqnarray}
On the other hand, the flow from a permutation brane and its
anti-brane to a tensor product brane is {\it non-perturbative}. 
In fact, the $g$-factors of the permutation branes approach 
\be
g(|\!|[L,M]\rangle\!\rangle) = \frac{1}{\sqrt{2}}\, 
\frac{\sin(\pi (L+1)/(k+2))}{\sin(\pi/(k+2))} 
\stackrel{k\rightarrow \infty}{\longrightarrow} 
\frac{L+1}{\sqrt{2}} \,,
\ee
whereas 
\be
\lim_{k\rightarrow\infty} g(|\!|L_1,L_2\rangle\!\rangle) = 0 \,.
\ee
Thus the ratio of the $g$-factors is zero in the limit 
$k\rightarrow \infty$. Such non-perturbative flows do not 
necessarily preserve the $K$-theory charges of the corresponding
branes, and this is in fact what happens here. The permutation
branes carry RR-charge and therefore integer valued K-theory
charge. On the other hand, the tensor product branes carry only
torsion charge (as is familiar from the case of one minimal
model \cite{Fred,H}). The configuration of permutation branes that can
flow to a tensor product brane does not carry any RR (or indeed
K-theory) charge, and thus if the flow were to preserve the K-theory
charge, it would follow that the tensor product branes would not carry
any K-theory charge at all. 

In fact, the situation is analogous to the case of the non-BPS
D0-brane of Type I theory \cite{Sen1}. This D-brane can be obtained
from the superposition of a D1-brane anti-D1-brane pair by taking the
tachyon to be the kink solution. The resulting configuration carries
non-trivial torsion $\Zop_2$ K-theory charge. On the other hand, the
original configuration of a D1-brane anti-D1-brane pair with a
constant tachyon solution carries trivial K-theory charge. Changing
the trivial tachyon solution to the kink therefore does not preserve
the K-theory charge (since the value of the tachyon is changed at
spacelike infinity). 

The fact that this flow does not preserve the K-theory charge is also
visible in the matrix factorisation description: the relevant
tachyonic operator that needs to be switched on in order to flow from
the superposition of permutation branes to the tensor product brane is
not the `constant' mode. [Indeed, as is explained in section~3.3.2,
$\lambda=x_1+\eta_n x_2$ in (\ref{lambdadef}).] Rather, it can be
thought of as being the analogue of the kink solution.

\section{The Gepner Model}

Now we would like to use the permutation branes analysed in the
previous sections as building blocks for branes in Gepner models (see
\cite{R} for the original construction of a generalised class of such
boundary states). This is the conformal field theory analog of the
constructions pursued in \cite{ADD,ADDF} from the matrix factorisation
side. 

The Gepner model is an orbifold of a tensor product of $N=2$ minimal
models whose central charges add up to $9$. We restrict
our discussion to the case of five minimal models. Together with a free
field theory with $c=3$ it describes a Calabi-Yau compactification in
the light cone gauge. As opposed to the earlier sections, we work in
the following with the minimal model before GSO projection. Its
Hilbert space is given by 
\beq
\H= \bigoplus_{[l,m,s]} \left[
\left(\H_{[l,m,s]} \otimes
\bar\H_{[l,m,s]}\right) \oplus \left(\H_{[l,m,s]} \otimes
\bar\H_{[l,m,s+2]}\right)
\right]\,.
\eeq
The Hilbert space of the Gepner model before orbifolding is the tensor
product of five such models, subject to the constraint that the
world-sheet spin structures of the minimal models are properly
aligned, {\it i.e.} that only states for which the $s_i$ are all
even, or the $s_i$ are all odd appear. We denote the generator of the
$\ZZ_{k_i+2}$ axial symmetry (\ref{gacdef}) in the $i$th minimal model
by $g_{i}$. The generator of the Gepner orbifold is then 
$g=g_1\cdots g_5$; its order is $H={\rm lcm} \{k_i+2 \}$. In the
$n$th twisted sector (where $n=1,\ldots, H-1$), the right-moving
$\bar{m}_i$ differ then from the left-moving $m_i$ by $2n$ (for all
$i=1,\ldots,5$); see \cite{RS} for explicit expressions for the
partition function and \cite{FSW} for a detailed  discussion of the
necessary projections.  
\smallskip

We want to construct boundary states that involve the permutation
boundary states for the tensor product of two minimal models.
To do so, we assume that $k\equiv k_4=k_5$. Our construction closely
follows the discussion of RS boundary states \cite{RS} given in
\cite{BHHW}, to which we refer for further details including also
short orbit states \cite{FSW}, which we will not discuss here. 
The original construction of permutation boundary states 
for more general permutation groups in Gepner
models appeared previously in \cite{R}.

In each sector (NS-NS and R-R and for each spin structure $\eta$) we
will consider a boundary state that is a tensor product of a standard
(Neumann or Dirichlet) free field boundary state (corresponding to the
free $c=3$ theory), as well as a boundary state of the internal Gepner
theory. GSO-invariance in the closed string requires that we add 
together states with different spin structures, while in order to
obtain a GSO-invariant open string spectrum one needs to add NS-NS and
R-R sector components. (For a review of these matters see for example
\cite{Gabrev}.) The GSO-projection however only applies to the full
ten-dimensional theory; thus the sum over the different
spin-structures can only be done once all components have been
tensored together.

In the following we shall only consider the internal part of each such
constituent boundary state. As we have seen before, the spin structure
$\eta$ is related to whether the label $S_i$ is even or
odd; in the following we shall therefore always consider the case
where either all $S_i$ are even, or all $S_i$ are odd. Furthermore, we
choose the convention that NS-NS components are labelled by $s=0$ and
R-R components by $s=1$. The constituent states for the usual and the
permutation gluing conditions are described in detail in appendix~D.

\subsection{The transposition branes}

On the basis of these expressions it is then straightforward to write
down a Gepner boundary state that consists of tensor product boundary
states in the first three factors, and a permutation boundary state
for the last two:
\begin{eqnarray}\label{gepnerstate}
&|\!|& L_1,L_2,L_3,L,M,\hat{M},S
\rangle\!\rangle_{(-1)^{(s+1)F}} \\
& & =
\frac{1}{\sqrt{H}}\sum_{n \in \Zop_H} \, 
\sum_{l_1,l_2,l_3,l,m}\, \sum_{\nu_i\in\Zop_2}\, 
e^{-\pi i \frac{\hat{M}n}{H}} (-1)^{S\sum \nu_i} 
\, \prod_{i=1}^3 \left((2k_i+4)^{\frac{1}{4}}
\frac{S_{L_i l_i}}{\sqrt{S_{0l_i}}} \right) \,
\frac{1}{2} e^{\pi i \frac{Mm}{k_4+2}} \frac{S_{Ll}}{S_{0l}} \nonumber
\\
& & \qquad \qquad \times e^{-\pi i \frac{s}{2} S}\, 
|l_1,n,s+2\nu_1;l_2,n,s+2\nu_2;l_3,n,s+2\nu_3;
\nonumber \\
& & \qquad \qquad \qquad \qquad \qquad \qquad  \qquad \qquad 
[l,m+n,s+2\nu_4]\otimes
[l,-m+n,s+2\nu_5]\rangle\!\rangle \,. \nonumber 
\end{eqnarray}
Here we have summed over the contributions of the twisted sectors
(that are labelled by $n$), but not over the spin-structures, nor the
NS-NS and R-R sectors --- these sums can only be done once the
space-time part of the boundary states has also been included. 

Nevertheless, everything that is of interest can already be read off
from this expression. In particular, using the formulae from
appendix~D (in particular (\ref{Ct}) and (\ref{Cp})), the one-loop
amplitude becomes  
\begin{eqnarray} \label{permperm}
&&
\langle\!\langle L_1',L_2',L_3',L',M',\hat{M}',S'|\!|
q^{L_0+\bar{L}_0 -\frac{c}{12}}
|\!|L_1,L_2,L_3,L,M,\hat{M},S\rangle\!\rangle_{(-1)^{(s+1)F}}  \\
& & \quad = \frac{1}{2} \sum_{l_i,m_i,s_i} 
\delta^{(H)} \left(\frac{\hat{M}'-\hat{M}}{2} 
+ H\sum_{i=1}^{5} \frac{m_i}{2k_i+4}
\right) \, 
\prod_{i=1}^5 \delta^{(2)}(S-S'+s_i) \, 
e^{-\frac{\pi i s}{2}(S-S'+\sum s_i)} \nonumber \\
&& \qquad \quad \times \ \prod_{i=1}^3 N_{L_i'L_i}^{l_i} \, 
\sum_l N_{LL'}^l N_{l_4 l_5}^l \delta^{(2k+4)}(M-M'+m_4-m_5) \,
\prod_{i=1}^5 \chi_{[l_i,m_i,s_i]}(\tilde{q}) \,. \nonumber
\end{eqnarray}
Here the sums run over all quintuples of triplets $(l_i,m_i,s_i)$ such 
that $l_i+m_i+s_i$ is even. [The factor of $1/2$ accounts, as in 
(\ref{Cp}), for the fact that the equivalent representations
$(l_4,m_4,s_4), (l_5,m_5,s_5)$ and 
$(k-l_4,m_4+k+2,s_4+2), (k-l_5,m_5+k+2,s_5+2)$ appear twice in the
above expression.]
 
The permutation boundary states preserve one half of the
space time supersymmetry, the phase of which is determined by the
label $\hat{M}$, just like for the ordinary tensor product branes.
Note that one can explicitly confirm from the one-loop amplitude
that the spectrum on every brane is tachyon-free (once the final
GSO-projection has been performed); in fact, this is simply a
consequence of the fact that (\ref{permperm}) depends in the usual 
manner on $s (S-S'+\sum_i s_i)$. Hence, the branes are stable \cite{R}.  
\medskip

In order to determine the charge lattice spanned by these 
boundary states, it is of great interest to calculate the open string 
Witten index $\Tr_R (-1)^F$ between the branes with different
$M$ and $\hat{M}$. In order to isolate this contribution from 
the above overlaps, one has to take the R-R component
({\it i.e.}\ the $s=1$ component of (\ref{permperm})), and consider
the overlap between the boundary states with opposite spin 
structure. [Recall that in the full boundary state one has to add
to the above boundary state in each sector its image under
$(-1)^{F_L}$ in order to make it invariant under the closed string
GSO-projection.] The action of $(-1)^{F_L}$ on each boundary state
shifts each $S_i$ by one, but also shifts $\hat{M}$ by $-H$. (This is
the case both for tensor product boundary states, as well as for
permutation boundary states.) Taking all this into account, one
obtains 
\beqa
&& I(L_1',L_2',L_3',L',M',\hat{M}',S'|L_1,L_2,L_3,L,M, \hat{M},S) 
\nonumber \\ 
&& \qquad = - \sum_{m_i} \delta^{(2k_4+4)}(M-M'+m_4-m_5) \, 
\, \delta^{(H)} 
\left(\frac{\hat{M}'-\hat{M}+H}{2}
+H \sum_{i=1}^5 \frac{m_i}{2k_i+4}\right) \nonumber \\
& & \qquad \qquad \qquad \qquad \qquad \times 
\prod_{i=1}^3 \hat{N}_{L_i'L_i}^{m_i-1}\sum_l N_{LL'}^{l}
N_{m_4-1, m_5-1}^l e^{-\frac{\pi i}{2}(S-S')} \,, 
\eeqa
where $\hat{N}$ denotes the periodically continued fusion rule
coefficients. Of particular interest is the basic case 
$L_i'=L_i=L=L'=0$, for which the above formula simplifies further
\beqa\label{45CFT}
&& I(0,0,0,0,M',\hat{M}',S'|0,0,0,0,M,\hat{M},S) \nonumber \\
& &  \qquad = - \sum_{m_i} 
\delta^{(2k_4+4)}(M-M'+m_4-m_5) \, \delta^{(H)} 
\left(\frac{\hat{M}'-\hat{M}+H}{2}
+H \sum_{i=1}^5 \frac{m_i}{2k_i+4}\right) \nonumber \\
& & \qquad \qquad \qquad \qquad \qquad \times
\left(\prod_{i=1}^3 \hat{N}_{00}^{m_i-1} \right)\,
N_{m_4-1,m_5-1}^0 \, 
e^{-\frac{\pi i}{2}(S-S')} \,.
\eeqa
This index is sometimes independent of the labels $M,M'$; in
particular this is the case for $w_4=w_5=1$. 
Along the lines of \cite{BDLR,DR} one can then rewrite this intersection
number in terms of the symmetry generator $G$ that shifts $\hat{M}$
by $2$ ($G$ acts as the $H$-dimensional shift matrix). 
To this end one replaces the fusion rule coefficient in each  
factor by $(1-G^{-w_i})$, where $w_i$ is the weight
$w_i=H/(k_i+2)$, and accounts for the $\delta^{(H)}$-function
constraint by some additional overall factor. 
For $w_4=w_5=1$ one obtains for the index\footnote{For the quintic this
formula was already found in \cite{R}.}
\be\label{45}
I_{(45)}  = 
G^{-w_4}\, \prod_{i=1}^3 (1-G^{-w_i}) 
= - G^{w_4} \prod_{i=1}^3 (1-G^{w_i})\,.
\ee
It is easy to see that the same formula also holds in the case 
$w_4 \neq 1$, provided that $M'-M \neq 0$. 

In order to determine the charges of the permutation branes, it is
also important to determine the overlap of a permutation brane with
one of the tensor product branes. There is again a subtlety in the
calculation of the overlaps between the corresponding Ishibashi
states; the correct generalisation of (\ref{import}) that is invariant
under field identifications is now (we are again restricting ourselves
just to two factors)
\begin{eqnarray}
\langle \!\langle l,n,s;l,n,s | q^{L_0+\bar{L}_0 - \frac{c}{24}} 
| [l,n,s]\otimes [l,n,s]\rangle\!\rangle & = & 
\hbox{Tr}_{[l,n,s]\otimes[l,n,s]} (\sigma \, q^{L_0-\frac{c}{24}} ) 
\nonumber \\ \label{import1}
& = & e^{\frac{\pi i n}{k+2} -\frac{\pi i s}{2}} \chi_{[l,n,s]}(q^2)
\,.  
\end{eqnarray}
The phase factor amounts to an insertion of $(-1)^{F_L}$, which takes
into account the statistics of the states that are permuted by
$\sigma$. 

Then it is straightforward to calculate the overlaps between tensor
product and permutation branes in the Gepner model, and one finds 
\begin{eqnarray}\label{permtensor}
&& \langle\!\langle L_1', \ldots, L_5',\hat{M}',S'|\!|
q^{L_0+\bar{L}_0-\frac{c}{24} }
|\!| L_1, L_2, L_3,L, M, \hat{M}, S 
\rangle\!\rangle_{(-1)^{F(s+1)}} 
\\
& & \quad =\sum_{l_i,m_i,s_i} \delta^{(H)} 
\left(\frac{\hat{M}'-\hat{M}+1}{2} 
+ H \sum_{i=1}^4 \frac{m_i}{2k_i+4}\right) \,
\prod_{i=1}^3 \delta^{(2)}({S-S'-s_i}) \,\delta^{(2)} ({s_4+1}) 
\nonumber \\
& & \qquad \quad \times 
e^{-\frac{\pi i s}{2} (S-S'+\sum_{i=1}^4 s_i + 1)} \,
\prod_{i=1}^3 N_{L_i'L_i}^{l_i} 
\sum_{\hat{l}} N_{L_4'L_5'}^{\hat{l}} N_{Ll_4}^{\hat{l}} \,
\prod_{i=1}^3 \chi_{[l_i,m_i,s_i]}(\tilde{q}) \,
\chi_{[l_4,m_4,s_4]}(\tilde{q}^{\frac{1}{2}})\,. \nonumber 
\end{eqnarray}
This overlap is in particular independent of $M$. 

As before, the index is simply the open string Witten index
$\Tr_R(-1)^F$. This can be calculated in the same manner as the index
between two permutation branes, and one finds
\beqa
&&I(L_1', \dots L_5', \hat{M}',S'|L_1,L_2,L_3,L,M,\hat{M},S) \nonumber
\\ 
& &  \qquad \qquad \qquad = - \sum_{m_i} \delta^{(H)} 
\left(\frac{\hat{M}'-\hat{M}+1+H}{2}
+H \sum_{i=1}^4 \frac{m_i}{2k_i+4}\right) \\
& &  \qquad \qquad \qquad \qquad \qquad \times
e^{-\frac{\pi i}{2} (S-S')}
\prod_{i=1}^3 \hat{N}_{L_i'L_i}^{m_i-1}
\sum_{\hat{l}} N_{L_4'L_5'}^{\hat{l}} \hat{N}_{Lm_4-1}^{\hat{l}}\,,
\nonumber 
\eeqa
where $\hat{N}$ are the periodically continued fusion rule
coefficients. Since the charges of all tensor product and all
permutation boundary states can be obtained by forming bound states of
$L=0$ branes, we are particularly interested in the index for the
brane with $L_i=L_i'=L=0$, in which case the above formula simplifies
to 
\beqa\label{intersect}
&& I(0,0,0,0,0,\hat{M}',S'|0,0,0,0,M,\hat{M},S) \nonumber \\
& & \qquad \qquad \qquad = - \sum_{m_i} 
\delta^{(H)} \left(\frac{\hat{M}'-\hat{M}+1+H}{2} 
+ H \sum_{i=1}^4 \frac{m_i}{2k_i+4}\right) \\
& &  \qquad \qquad \qquad \qquad \qquad \times
e^{-\frac{\pi i}{2} (S-S')}
\left(\prod_{i=1}^3 \hat{N}_{00}^{m_i-1} \right)
\hat{N}_{0,m_4-1}^0\,. \nonumber 
\eeqa
Rewriting this in terms of the symmetry operator $G$ leads to 
\beq
I_{(45)-RS}= - \prod_{i=1}^{4} (1 - G^{-w_i}) = 
- G^{w_4}\, \prod_{i=1}^4 (1-G^{w_i}) \,.
\eeq

\subsection{The (23)(45) permutation branes}

Until now we have only considered the permutation branes that involve
a single transposition in the last two factors, and usual B-type
boundary states for the first three factors. Using the same
ingredients we can obviously also construct the boundary states where
we combine two transpositions in the factors $k_2=k_3$ and $k_4=k_5$, 
with a usual B-type boundary state for the first factor.
The calculations follow the same pattern as above, in particular, one
just needs to collect the results of the building blocks summarised in
appendix~D. As before, the resulting branes are also stable.

\noindent The intersection matrix between two such branes has a simple
form if $w_2=w_4=1$ or if both $M_1\ne M_1'$ and $M_2\ne M_2'$, in
which case it is (for the quintic this formula was already given in
\cite{R})  
\beq\label{PP}
I_{(23)(45)}= (1-G^{w_1}) \, G^{w_2}\, G^{w_4} \,.
\eeq
For the intersection between the RS and these permutation branes we
obtain on the other hand
\beq\label{(23)(45)-RS}
I_{(23)(45)-RS}=(1-G^{w_1})\, (1-G^{w_2})\, (1-G^{w_4})
\, G^{w_2} \, G^{w_4} \,.
\eeq
Finally, the intersection between the (45) and the (23)(45) boundary
states is 
\beq\label{(23)(45)-(45)}
I_{(23)(45)-(45)}=(1-G^{w_1})\, (1-G^{w_2})\, G^{w_2} \, G^{w_4}
\,. 
\eeq
This last formula is again in general only correct if $w_4=1$ or the
$M$-labels corresponding to the (45) permutation of the two branes are 
different.

\section{Gepner model, matrix factorisations and geometry}

At this point it is natural to ask what the large volume 
charges of these permutation branes are. For the RS branes, this 
question has first been answered in several examples in 
\cite{BDLR,DR,KLLW,SchI}, using the analytic continuation of periods
to the Gepner point. In later work \cite{DD,T,Mayr,DFR,GJ} it has been
understood that the RS branes correspond geometrically to pull backs
of certain bundles from the embedding projective space to the
Calabi-Yau hypersurface. 

The lattice generated by the RS branes is only a sublattice of the
full charge lattice which is only in special cases of maximal rank. It
also typically does not contain the charge vectors of minimal 
length. One may wonder whether the permutation boundary states may
give rise to new charges and to charge vectors of minimal length;
as we shall see in this section, both phenomena occur. In fact, at
least for a number of examples (including the case of the quintic), a 
certain family of permutation branes can be shown to span the full
charge lattice at the Gepner point. However, we also show, that there
are examples where the permutation branes do not account for all
charges.

\subsection{The transposition branes}

Using the results of \cite{DD,T,Mayr,DFR,GJ} one can calculate the 
large volume charges of the RS branes in generic models. Whenever the
transposition and the RS branes generate the same subspace of the RR 
charge lattice (this is in particular the case when $w_4=w_5=1$) we
can determine the charges of the transposition branes via the
intersection matrix between permutation and tensor product branes. Our
analysis of the flows between rank $1$ factorisations and tensor
product factorisations in section~3.3.2 then suggests that the
RS branes can be obtained as bound states of the transposition branes,
and thus that the RS charges are linear combinations (with integer 
coefficients) of the charges of the transposition branes. In fact, it
is clear from the above equations that (at least for $w_4=w_5=1$)
\be
I_{(45)} \, (1 - G^{w_4}) = I_{(45) - RS} \,.
\ee
Thus the integer matrix $(1 - G^{w_4})$ expresses the charges of the
RS branes in terms of those of the (45)-branes. This relation is also
consistent with the relation between the intersection forms 
\beq
I_{RS} = (1-G^{w_4})^t\, I_{(45)} \, (1-G^{w_4}) \,,
\eeq
where $I_{RS}$ is the intersection form of the RS branes which, in our
conventions, equals 
\be
I_{RS} =  \prod_{i=1}^{5} (1 - G^{w_i}) \,.
\ee
It also agrees with the intersection matrix involving the (23)(45)
branes (if they exist),
\be
I_{(23)(45)-(45)} (1 - G^{w_4}) = I_{(23)(45)-RS} \,.
\ee

It is remarkable that the intersection forms calculated above
for the permutation branes (where the permutation is applied to two
models with $w=1$ in the Gepner model)
exactly coincide with the `intersection matrix in the Gepner basis'
computed for one- and two-parameter 
examples in \cite{BDLR,DR,KLLW,SchI}.\footnote{For the case of the
quintic this was also noted before in \cite{R}.} In those papers, the 
Gepner basis was taken from \cite{CDGP,F,KT,CDFKM,CFKM}, where it was 
obtained by analytic continuation of the fundamental
period at large volume. The fundamental period is the solution of
the Picard-Fuchs equations without logarithms, and corresponds to
a D0 brane. Subsequently, the $\ZZ_H$ Gepner monodromy transformation
was applied to this period to obtain $H$ periods (with linear
relations) at the Gepner point. By construction, the Gepner monodromy
takes a particularly simple form in this basis, since it is just an
$H$-dimensional shift matrix.  

The transformation between the Gepner basis and the large volume basis 
was given in the old papers  \cite{CDGP,F,KT,CDFKM,CFKM}; in
particular, the fundamental period is directly mapped to one of the
periods in the Gepner basis. Historically, this transformation,
together with the relation between Gepner and RS basis provided the
first way to determine the charges of RS branes at large volume. 

Here, the relation between the RS basis and the Gepner basis 
is given by the same change of basis as between (45)-
and RS branes. Thus we can conclude that one of the (45)-branes
corresponds to the D0 brane, or the fundamental period, that has 
been used in the original construction of the Gepner basis. For the
quintic, the relation between the (45)-boundary state and D0-branes
was already mentioned in \cite{R1}, where it was however claimed to
correspond to $5$ D0-branes. 

Table 1 gives an overview over the models where the
Gepner intersection matrix was obtained by analytic continuation
and shown by our analysis to coincide with the single permutation
intersection matrices $I_{(45)}$, where two models with $w=1$ 
are permuted. [To be consistent with the labelling of the models
below, these branes are more appropriately referred to as (12)-branes!]
Hence, we know that D0 branes exist as rational boundary states in all
of these models. To check this, we have verified by explicit counting
that all of the relevant boundary states possess $3$ candidate
marginal operators. [In the case that the Gepner model consists of
only $4$ factors, one has to add into the formula for the intersection
matrix a factor of $(1-G^{H/2})$ --- see \cite{FKLLSW} for a
discussion of the difference in the open string projection.]  
$$
\begin{array}{|c|c|c|c|}
\hline
{\rm CY-hypersurface} & {\rm Gepner}\,{\rm model} & I_{(12)} & 
{\rm Reference} \\ \hline
\hline
{\IP}_4[5] & (k=3)^5 & - G (1-G)^3 & \cite{CDGP,BDLR,R}\\
\hline
\IP_{(1,1,1,1,2)}[6] & (k=4)^4 (k=1)^1 & 
-G (1-G^{1})^2 (1-G^{2}) & 
\cite{KT,F,SchI}\\
\hline
\IP_{(1,1,1,1,4)}[8] &  (k=6)^4 & 
- G  (1-G^{1})^2 (1-G^{4})  & 
\cite{KT,F,SchI}\\
\hline
\IP_{(1,1,1,2,5)}[10] & (k=8)^3 (k=3) &
- G (1-G) (1-G^2) (1-G^{5}) &  
\cite{KT,F,SchI} \\
\hline
\IP_{(1,1,1,6,9)}[18] & (k=16)^3 (k=1) & 
- G (1-G) (1-G^{6})(1-G^{9}) & 
\cite{CFKM,DR} \\ 
\hline
\IP_{(1,1,2,2,2)}[8] & (k=6)^2 (k=2)^3 & 
- G (1-G^{2})^3  & 
\cite{CDFKM,KLLW,SchI} \\
\hline
\IP_{(1,1,2,2,6)}[12] & (k=10)^2 (k=4)^2 & 
-G (1-G^{2})^2 (1-G^{6}) & 
\cite{CDFKM,KLLW,SchI} \\
\hline
\end{array}
$$

\subsubsection{Stability}

By constructing an explicit, stable boundary state of the right
charges, we have therefore also settled the question of whether the D0
brane on the quintic (and several other models) is stable at the
Gepner point.  This is in agreement with the analysis of the recent
paper \cite{Wal}, where a stability condition based on matrix
factorisation was formulated and applied to the case of the
quintic. The existence of  D0 branes in the stringy regime has been
addressed in the literature from various points of view. In
\cite{BDLR} it was noted that the position of a D0 brane would break
part of the $\ZZ_5^4$ symmetry of the quintic. On the other hand,
since all RS branes are invariant under this symmetry, it could  not
be expected to find a D0 brane among them. Rational RS boundary states
carrying only (multiple units of) D0 brane charge are known to exist
in other models, where this argument fails; a systematic search was
performed in \cite{Sch}. Since the permutation boundary states we
constructed transform non-trivially under the $\ZZ_5$ symmetry scaling
$x_4$, the symmetry is explicitly broken so that the above arguments
do not apply. 

In a wider context, the stability of D-branes depending on 
K\"ahler moduli has been addressed in the context of 
$\Pi$-stability \cite{DFR,Pi}. D-branes on the quintic, and in particular
D0 branes, were studied in \cite{AD}, who plotted the lines of
marginal stability in several cases. (See also \cite{AL} for an
analysis of D0 brane stability.)  Their analysis shows that there
exists a line of marginal stability where the D0 brane is destabilised
by the D0-D6 system (that would be unstable in the large volume
regime). The reason is, roughly speaking, that at the conifold point
the period associated to the D6 brane shrinks to zero, so that it
becomes preferable for the brane to wrap that period. The D0 brane is
however stable against a decay into a D0-D6 + anti-D6 system in the
stringy regime, providing a strong argument for the existence of D0
branes at the Gepner point.

\subsubsection{Geometry from matrix factorisation}

The interpretation of one of these boundary states as a D0-brane is
also strongly suggested by the description in terms of matrix
factorisations. Geometrically, D0-branes are points on the Calabi-Yau 
and can be described by a set of $4$ linear equations
$K_1=K_2=K_3=K_4=0$, together with the defining 
equation of the Calabi-Yau hypersurface 
$W=x_1^{w_1}+x_2^{w_2}+x_3^{w_3}+x_4^{w_4}+ x_5^{w_5}$.\footnote{Note
however that in the case that the vanishing locus of the $4$ linear
equations intersects with an exceptional divisor, the brane will be
higher dimensional; we will encounter an example of this below. If
$w_4=w_5=1$, this is excluded.} 

To obtain a matrix factorisation related to these equations, one can
proceed as in \cite{ADD} (where the case of the quintic 
was treated) and find four polynomials $F_1,\dots, F_4$ such that  
\be
\sum_{i=1}^4 K_i F_i=W \,.
\ee
Given this relation one can immediately write down a matrix
factorisation, iterating the step reviewed around equation
(\ref{tensorfac}). Using the relation between matrix factorisations
and the geometrical category of singularities \cite{O} reviewed in
section~3.1, one can conclude, as in \cite{ADD}, that such a
factorisation is a good candidate for a D0 brane on the Calabi-Yau
hypersurface. The location of the D0 brane is described by the zero
set of the four linear equations.
In particular, to make contact with our previous
discussion on the relation between permutation boundary states and
matrix factorisations, one can  consider the special case where   
\beq\label{D0fac}
K_1 =x_1\,, \quad K_2=x_2\,, \quad K_3=x_3\,, \quad
K_4= x_4-\eta x_5 \,,
\eeq
which geometrically singles out a set of points on the Calabi-Yau
hypersurface. For the case of the quintic, it is shown in \cite{ADD}
that the deformation space of these objects is parametrised by points
on the quintic, as it should be for a D0-brane. 

Our detailed comparison between boundary states in the tensor product
of two minimal models and matrix factorisations immediately suggests
that the boundary state in the Gepner model that corresponds to this 
factorisation is given by the tensor product of three minimal
model branes with a permutation brane in the last two factors. The
labels of the boundary state should be $L_i=0$, $L=0$, while the choice
of $M$ corresponds to the choice of $\eta$ in (\ref{D0fac}). This is
precisely the boundary state we considered above.

The charge of the brane can be verified on the matrix side by the
same type of index calculation that we performed above using conformal
field theory methods. Indeed, given our detailed comparison between
matrix factorisations and conformal field theory it is
clear that the two calculations lead to the same result, see
\cite{ADD} for the quintic.
\medskip

For two of the above examples, namely $\IP_{(1,1,1,2,5)}[10]$ and 
$\IP_{(1,1,1,6,9)}[18]$, one can easily show that the intersection
matrix of the D0-brane and its images under the Gepner monodromy span
already the full charge lattice. 
Indeed, for  $\IP_{(1,1,1,2,5)}[10]$ ($\IP_{(1,1,1,6,9)}[18]$) the
intersection matrix
contains a 4-dimensional (6-dimensional) submatrix of determinant $1$.
In the other cases, however, the relevant submatrices
of maximal rank have determinant bigger than $1$. In these cases there
exists a second construction which we will discuss in the next section.

To conclude this section, let us discuss the two-parameter example
$\IP_{(1,1,2,2,2)}[8]$ in more detail. We first summarise its
geometrical properties which have been discussed in \cite{CDFKM}, to
which we also refer for further details. The cohomology ring is
generated by the two divisor classes $L$ and $H$. $L$ corresponds to
the zero locus of degree one equations, whereas $H$ is the zero locus
of degree two equations. The two divisors intersect along a curve 
\beq 
4h=HL \,.
\eeq
The curve $x_1=x_2=0$ is a priori singular, and the singularity is
resolved by blowing up each point on it into a $\IP_1$ 
denoted by $l$, 
\beq
4l=H^2-2HL\,.
\eeq
Furthermore, we define
\beq
4v = H^3/2 =H^2L\,.
\eeq
There are two types of $(45)$ branes in this model: first, we can
consider the equations\footnote{This brane should be more
appropriately referred to as the $(12)$-brane.}
\setcounter{footnote}{0}
\be
x_1=\eta x_2, \quad x_3=x_4=x_5=0 \,.
\ee
As was already mentioned before, this is geometrically a D0-brane. The
second option is to consider
\be
x_1=x_2=x_3=0, \quad x_4=\eta x_5\,.
\ee
This describes geometrically a $\IP_1$ coming from the blow up of the
singular curve. Hence, the matrix considerations suggest that this
brane is a D2-brane. Indeed, calculating the brane charges via the 
intersections with the RS branes, one can confirm that the set of $8$
(45) branes contains a single brane wrapping this cycle.

\subsection{The $(23)(45)$-branes}

The second class of branes can be constructed whenever two pairs of
levels coincide. Their intersection matrix is given by
(\ref{PP}) (at least if $w_2=w_4=1$). Given (\ref{(23)(45)-(45)}), it
is again clear that whenever such (23)(45)-branes exist, the charges
of the (45)-branes can be expressed in terms of integer linear
combinations (that are determined by the matrix $(1-G^{w_2})$) of the
(23)(45)-branes. Thus the latter are always more fundamental.

We will now show that the intersection matrix (\ref{PP}) always
contains a submatrix of dimension $H-w_1$ which has determinant
$1$. First we observe that in order to determine the determinant of a
submatrix (up to a sign), we can ignore the factors of $G^{w_2+w_4}$,
and consider $I'=(1-G^{w_1})$ instead. It is easy
to see that this matrix has a kernel of dimension $w_1$; if we remove
the first $w_1$ rows and columns the resulting matrix is upper
triangular with $1$s on the diagonal. Hence this submatrix of
dimension $H-w_1$ has determinant $1$.  

This is already sufficient to show that the $(23)(45)$-branes generate
the full charge lattice in the remaining examples above.\footnote{We
have checked explicitly in these examples, that the relevant
intersection matrix is indeed of the form (\ref{PP}), although
$w_2=w_4=1$ does not hold.}  As before, this basis of the charge
lattice is particularly natural at the Gepner point, since the Gepner
monodromy just acts by a permutation. This feature is shared by the
Gepner bases given in the literature (that, as we have argued,
correspond to $(45)$-permutation boundary states); in general however,
the latter are not minimally normalised.  
\smallskip

As before, one can use the matrix description to propose a geometrical
interpretation of these branes, following \cite{ADDF} who considered
only the quintic. According to the general discussion, one is led to
propose that the boundary states correspond to the zero locus of the
linear equations
\be\label{D2define} 
K_1=x_1 =0\,, \quad K_2=x_2-\eta_2 x_3=0\,, 
\quad K_3=x_4-\eta_4 x_5=0\,. 
\ee
Generically, this system of equations describes a D2-brane.
As before, one can then find $F_i$ such that $W=\sum K_i F_i$,
leading to a matrix factorisation. 
For the case of the quintic, the deformation theory and superpotential
for these theories has been studied in \cite{ADDF}, and known 
geometrical results on the obstruction theory of lines on the quintic
have been reproduced.

In the case of the two-parameter model $\IP_{(1,1,2,2,2)}[8]$, 
(\ref{D2define}) defines the intersection of an
element of the divisor class $L$ with an element of the divisor class
$H$. The result should therefore be a D2 brane wrapping the cycle
$h$. This can again be verified by an explicit calculation via the
index. 

Since the $(23)(45)$ branes provide a minimal basis at the Gepner
point, we list their large volume charges in our two main
examples. Here, one uses the charges of the RS branes, which
correspond to the pure D6 brane and its monodromy images. The
labelling is chosen such that the D6 has label $5$. For the quintic,
one obtains accordingly $5$ branes whose Chern characters are 
\beqa
{\rm ch}(V_1) &=& 2-H -\frac{3}{10} H^2 +\frac{7}{30} H^3 \nonumber \\
{\rm ch}(V_2) &=& -1 +H -\frac{3}{10} H^2 - \frac{7}{30} H^3 \nonumber \\
{\rm ch}(V_3) &=& \frac{1}{5} H^2 - \frac{1}{5} H^3 \nonumber \\
{\rm ch}(V_4) &=& \frac{1}{5} H^2 \nonumber \\
{\rm ch}(V_5) &=& -1 +\frac{1}{5} H^2 + \frac{1}{5} H^3 \,,
\eeqa
where $H$ is the integral generator of $H^2(M,\ZZ)$ with $M$ being the
quintic. The pure D2-brane is described by $V_4$.

For the two parameter model $\IP_{(1,1,2,2,2)}[8]$ we can use
the list of Chern characters of the RS branes given in \cite{DD}.  
The cyclic ordering of the RS charges is such that the brane with label
$8$ is the pure D6-brane, while the others are its monodromy
images. We find $8$ branes whose Chern characters are given by 
\beqa
{\rm ch}(V_1)&=&-1+h+H-l-L-\frac{5}{3} v \nonumber \\
{\rm ch}(V_2)&=& -3h+L \nonumber \\
{\rm ch}(V_3)&=& 2-3h-H-l+L+ \frac{2}{3} v \nonumber \\
{\rm ch}(V_4) &=&h-L+v \nonumber \\
{\rm ch}(V_5) &=& -1+h+l+v \nonumber \\
{\rm ch}(V_6) &=&h  \nonumber \\
{\rm ch}(V_7) &=& h+l \nonumber \\
{\rm ch}(V_8) &=& h-v \,.
\eeqa
In this case $V_6$ is the pure D2-brane.

\subsection{New charges from permutation branes}

In general, the charge lattice spanned by the RS branes does not
contain all charges of the model. In fact, the RS branes preserve a
very large symmetry, and thus only relatively few Ishibashi states can
be used in the construction. Geometrically, the RS branes have been
identified with pullbacks of certain rigid bundles to the covering
space. In general, however the cohomology of the hypersurface can be
quite different from that of the embedding space; for example the
torus, which has a holomorphic one-form, can be embedded as a
hypersurface in $\IP_2$, which does not have a one-form.

In this section, we will study the example $\IP_{(1,1,1,3,3)}[9]$,
or, as a Gepner model, $(k=7)^3(k=1)^2$. In this example, the
permutation branes give rise to new charges, as one can see from
a calculation of the rank of the intersection matrices. The model
has $h^{(1,1)}=4$, such that we expect four D2 and four D4 branes.
Together with the D0 and D6 brane, they span a 10 dimensional charge
lattice. It was realised in \cite{BD} that the RS branes do not carry  
all of these charges, but that one can project out the missing ones
by taking a suitable free orbifold, leading to a two-parameter model
with torsion in K-theory. 

We can use the formulae of chapter~5 to verify that 
the intersection matrix of the (45)-branes, which in this case
equals (for $M=M'$)
\be\label{45p}
I_{(45)} = (1+G^6) (1-G)^3
\ee
has rank $8$, while that of the RS branes only has rank $6$.
Note that in this case $w_4=3$, so that the above intersection form
is not of the standard form (\ref{45}) but was  derived directly
from the conformal field theory formula (\ref{45CFT}). We have also
calculated the intersection matrix for the (12)(45) branes which turns
out to be 
\be
I_{(12)(45)} = - G (1+G^6) (1 - G)  \,,
\ee
and also has rank $8$. In order to account for the
full charge lattice one observes that the intersection matrix of the
(45) branes at $M=0$ and $M=2$ has actually rank $10$. [This
intersection matrix is of the form
\be
I_{(45);0,2} = \left(\begin{matrix}
(1+G^6) (1-G)^3 \qquad & - G^3 (1-G)^3 \cr
G^{-3} (1-G^{-1})^3 \qquad & (1+G^6) (1-G)^3 
\end{matrix}\right)
\,,
\ee
as follows from (\ref{45p}) as well as from (\ref{45}).] The same is
the case for the intersection matrix of the (12)(45) branes at
$M_1=M_2=0$ and $M_1=M_2=2$; in the latter case, this intersection
matrix contains a submatrix of determinant $1$. [The (45) branes only
generate a sublattice of index $9$.] Thus the full charge lattice is   
generated by the (12)(45) branes at $M_1=M_2=0$ and $M_1=M_2=2$ (and
all values of $\hat{M}$). 
\smallskip

The fact that the permutation branes generate extra charges fits nicely
with expectations from geometry and matrix factorisations as we will
now explain. Due to the divisibility properties of the weights,
the embedding projective space has a $\ZZ_3$ singularity that locally
looks like $\BC^3/\ZZ_3$. This singularity is resolved by an exceptional
$\IP_2$. The hypersurface intersects with the singular locus in the three
points $x_4^3+x_5^3=0$. These $3$ points are cut out and replaced by
exceptional divisors, leading altogether to $h^{(1,1)}=4$ for the
cohomology of the hypersurface \cite{BD}. We can single out the
singular points by the equations
\beq\label{4pardiv}
x_1=x_2=x_3=0\,, \quad x_4-\eta x_5=0 \,.
\eeq
One would then expect that the corresponding matrix factorisation,
constructed as in section~6.1 and 6.2, give a matrix description of
branes wrapping the exceptional divisors, or equivalently, that the
three $(45)$ transposition branes with different $M$ labels carry the 
relevant charges. This is in agreement with the calculation of the
rank of the intersection matrix mentioned above. On the other hand,
for fixed $M$ the rank of the intersection matrix is $8$, thus
accounting for the $2$ D4 charges (+ $2$ D2 +D0+D6) already carried 
by the RS branes, as well as one extra D4 (+ D2) brane charge.

As a further confirmation of this picture one can match the
symmetry transformations of the branes. Consider the action
\be
(x_1,x_2,x_3,x_4,x_5) \to (x_1,x_2,x_3,e^{\frac{2\pi i}{3}}x_4,
e^{\frac{4\pi i}{3}} x_5)\,.
\ee
This transforms the equation (\ref{4pardiv}) with different values for  
$\eta$ into each other, so that only their superposition remains
invariant. On the level of boundary states, one sees that the RS
states remain invariant under the corresponding action in conformal
field theory, whereas the label $M$ of the permutation branes gets
shifted as $M\to M+2$. This means that only the invariant combination
can carry a charge contained in the RS lattice. 
\smallskip

One may wonder whether permutation branes whose permutations involve
longer cycles could generate additional charges or describe other 
preferred bases. This does not seem to be the case. In fact, one can
easily see that the number of B-type Ishibashi states corresponding to
permutations with longer cycles is always smaller than that of a 
permutation whose longest cycles have length two. In this sense the
above constructions are already `optimal', and at least as far as
charges are concerned, it is sufficient to consider only
transpositions as we have done in this paper.

Finally one may wonder whether the permutation branes always generate
the full charge lattice. It is however easy to see that this is not
the case. 
A simple example is the manifold $\IP_{(3,3,4,6,8)}[24]$, 
or, as a Gepner model, $(k=6)^2 (k=4) (k=2) (k=1)$. This theory has
only one class of permutation branes (the (12)-branes), and one can
easily show that they generate a sublattice of rank $12$. On the other
hand, $h^{1,1}=7$ \cite{CLS,LS1}, and thus the full charge lattice has
dimension $16$. In fact, one can identify certain RR ground states of
this theory which are part of the even cohomology charge lattice, but
which cannot couple to {\it any} standard tensor product or
permutation branes.\footnote{We thank Stefan Fredenhagen for helping
us find this example and check this property.} 
For this theory, these symmetric constructions therefore do not
account for all the charges. The situation is therefore similar to the
case of WZW models of groups of higher rank (see for example
\cite{GGR1,GGR2}).   

{}From the matrix factorisation point of view, one can guess how to
obtain the remaining constructions --- these correspond probably to 
factorisations that rely on the fact that other pairs of $w_i$ have
common factors. This is also what one expects from geometry, since 
common factors of the weights lead to additional exceptional sets.
At this stage it is however not clear what the corresponding boundary
states in conformal field theory should be.

\section{Conclusion}

In this paper we have identified the D0-brane and D2-brane for a
number of Gepner models with certain permutation boundary states
\cite{R}. In some examples these D-branes, together with their images
under the Gepner monodromy, form a basis for the full charge
lattice. We have also shown that the permutation branes sometimes
carry charges that are not already accounted for by the RS branes. In
general, however, the permutation branes are not yet sufficient to
describe all the charges. 

Our analysis was inspired by the identification of the D0- and
D2-brane for the quintic in terms of matrix factorisations that was
given in  \cite{ADD,ADDF}. In particular, we studied the dictionary
between factorisations of $W=x_1^d+x_2^d$ with boundary states of the
tensor product of two minimal models. This allowed us to identify the 
relevant factorisations with permutation boundary states, and thus to
identify the boundary states of the D0- and the D2-brane. We also 
checked that these boundary states have the correct charges by
determining the Witten index with the RS boundary states whose charges
had been known before. (For the case of the quintic, this had also
been done, from the matrix point of view, in \cite{ADD,ADDF}.) Our
analysis shows in particular, that the D0-brane (as well as the
D2-brane) is stable at the Gepner point.    

In the examples we considered it was possible to predict the charge
of one out of the $H$ monodromy images from the form of the
factorisation, which indicated the location of the brane as the zero
set of linear equations. It would be interesting to generalise this,
including the effect of the GSO projection (choice of grading in the
matrix factorisation). This could presumably be done in the framework
of the linear sigma model where one can interpolate between large and
small volume, generalising the analysis for fractional branes in 
\cite{DD,T,Mayr,GJ}. 

As is explained in detail in section~4, we were only able to identify
a subset of rank $1$ factorisations with permutation boundary states;
it would clearly be interesting to understand the boundary state
description of the remaining factorisations. For the theories where
the permutation branes do not account for the full charge lattice, it
would also be interesting to understand (both from the matrix
factorisation point of view as well as in conformal field theory) the
branes that generate the remaining charges.

Concerning the relation between conformal field theory and matrix
factorisations, it would be interesting to understand conceptually the
relation between the various factorisations and their boundary states;
in particular, one may hope to be able to read off the symmetries that
are preserved by the brane ({\it i.e.} in particular the gluing
condition) from the structure of the factorisation.  Among other
things, this should clarify how the factorisations that correspond to
the permutation branes are singled out, and lead to clues for how to 
find the boundary state description of the remaining factorisations. A
further example, where a better understanding of symmetries might be
useful, are higher order permutation branes and their relation to
matrix factorisations.

\bigskip

\centerline{\large \bf Acknowledgements}
\vskip .2cm

This research has been  partially supported by a TH-grant 
from ETH Zurich, the Swiss National Science Foundation and the Marie
Curie network `Constituents, Fundamental Forces and Symmetries of the
Universe' (MRTN-CT-2004-005104). MRG is grateful to the Fields
Institute for hospitality during the final stages of this work. We
thank Stefan Fredenhagen, Davide Gaiotto, Terry Gannon, Manfred
Herbst, Kentaro Hori, Wolfgang Lerche, Andreas Recknagel and Daniel
Roggenkamp for useful discussions.

\appendix

\section{Conventions}

In this appendix we collect our conventions for the description of the
$N=2$ minimal models. The $N=2$ algebra is generated by the modes
$L_n$, $J_n$, $G^\pm_r$, subject to the commutation relations
\ba
{}[L_m,L_n] & = & (m-n) \, L_{m+n} + \frac{c}{12}\, m\, (m^2-1)\,
\delta_{m,-n} \nonumber \\
{}[L_m,J_n] & = & - n \, J_{m+n} \nonumber \\
{}[L_m,G^\pm_r] & = & \left(\frac{m}{2}- r\right) \, G^\pm_{m+r}
\nonumber \\ 
{}[J_m,G^\pm_r] & = & \pm\, G^\pm_{m+r} \nonumber \\
{}[J_m,J_n] & = & \frac{c}{3}\, m\, \delta_{m,-n} \nonumber \\
\{G^+_r,G^-_s\} & = & 2\, L_{r+s} + (r-s)\, J_{r+s} +
\frac{c}{3} \left(r^2 - \frac{1}{4}\right)\, \delta_{r,-s} \,.
\nonumber
\ea
Here $m$ and $n$ are integer; $r$ is integer in the R-sector, and 
of the form $r=\Zop+\frac{1}{2}$ in the NS-sector. The $N=2$ minimal
models occur for  
\be\label{central}
c = \frac{3 k}{k+2}\,,
\end{equation}
where $k$ is a positive integer. At least in this case, the bosonic
subalgebra of the $N=2$ algebra can be described in terms of the coset 
\be\label{coset}
\left(N=2\right)_{\ \rm bos} = 
\frac{su(2)_k \oplus u(1)_4}{u(1)_{2k+4}} \,.
\ee
Here $u(1)_d$ describes the U(1) theory whose representations are
labelled by integers mod $d$. The central charge of (\ref{coset})
obviously agrees with (\ref{central}).

The representations of the coset algebra are labelled by $(l,m,s)$,
where $l=2j$ with $j$ the (half-integer valued) spin of $su(2)$,
$m\in\Zop_{2k+4}$ and $s\in\Zop_4$. Since the $su(2)$ affine algebra
appears at level $k$, $l$ takes the values $l=0,1,\ldots, k$. These
labels are subject to the selection rule $l+m+s=0$ mod $2$. 
Furthermore we have the field identification 
\be
(l,m,s) \sim (k-l,m+k+2,s+2) \,.
\ee
The corresponding equivalence class will be denoted by $[l,m,s]$. We
shall usually suppress the level $k$ in our notation; also the central
charge $c$ will always be defined by (\ref{central}).

Since the coset algebra only describes the bosonic subalgebra of the
$N=2$ algebra, the irreducible representations of the $N=2$ algebra
consist of direct sums of representations of the coset algebra. In
fact, the NS-representations of the $N=2$ algebra correspond to the
sums 
\be
(l,m) = (l,m,0) \oplus (l,m,2)  \,,
\ee
where $l+m$ is even, and $(l,m)\sim (k-l,m+k+2)$. The
R-representations of the $N=2$ algebra correspond on the other hand to 
\be
(l,m) = (l,m,1) \oplus (l,m,3) \,,
\ee
where $l+m$ is odd, and $(l,m)\sim (k-l,m+k+2)$. In either case, we
also denote the corresponding equivalence class by $[l,m]$.

The conformal weights and the U(1)-charge (of the $N=2$ U(1)
generator) of the highest weight states of the coset representation
$(l,m,s)$ are, up to integers, given by 
\ba
h(l,m,s) & = & \frac{l(l+2) - m^2}{4 (k+2)} + \frac{s^2}{8} 
\\
q(l,m,s) & = & \frac{s}{2} - \frac{m}{k+2} \,. \label{uode}
\ea
In the NS sector, the chiral primary states appear in the
representations $(l,l,0)$ or $(l,-l-2,2)$. [Note that 
$(l,l,0)\sim(l,-l-2,2)$.] In the R sector, the condition for a chiral
primary state is $(l,l+1,1)$ or $(l,-l-1,-1)$. Finally, the modular
$S$-matrix of the coset theory is 
\be\label{Smatrix}
S_{LMS, lms} = S_{Ll}\, \frac{1}{\sqrt{2k+4}}\, 
e^{i\pi \frac{Mm}{k+2}}\, e^{-i\pi \frac{Ss}{2}} \,.
\ee
Here $S_{Ll}$ denotes the $S$-matrix of $su(2)_k$, which is explicitly
given as 
\be\label{Sma}
S_{Ll} = \sqrt{\frac{2}{k+2}}\, 
\sin\left(\pi\, \frac{(L+1)\, (l+1) }{k+2} \right)\,.
\ee
One easily checks that the $S$-matrix (\ref{Smatrix}) is unitary, and
that the above definition depends only on the equivalence class
$[L,M,S]$ and $[l,m,s]$. One also easily observes that 
\be\label{u1}
S_{LMS, lms+2} = (-1)^S\, S_{LMS,lms} \,.
\ee

\section{Open string spectra from matrix factorisations}

In this appendix we give details of the calculations to
determine the topological open string spectra between various
matrix factorisations of the product theory. 

\subsection{The open string spectrum between rank $1$ branes}

In the main part of the paper we only discussed the case where the two
branes in question are the same. Here we describe the calculation in
the general case, where
\be
J=\prod_{m\in I} (x_1-\eta_m x_2)\,, \qquad 
E = \prod_{n\in D\setminus I}
(x_1-\eta_{n} x_2) \,,
\ee
and
\be
\hat{J}=\prod_{\hat{m}\in \hat{I}} (x_1-\eta_{\hat{m}} x_2)\,, \qquad
\hat{E} = \prod_{\hat{n}\in D\setminus \hat{I}}
(x_1-\eta_{\hat{n}} x_2)\,.
\ee
We start with the discussion of the fermionic spectrum. Dividing the
first line of the BRST-invariance condition 
\ba\nonumber
0 &=& \hat{E}\, t_0 + t_1\, J \\
0 &=& \hat{J}\, t_1 + t_0\, E 
\ea
by the greatest common divisor of $\hat{E}$ and $J$, which is  
$\prod_{n\in I \setminus \{I \cap \hat{I}\}} (x_1-\eta_n x_2)$, we
obtain 
\be
t_0 = b(x_1,x_2) \prod_{n\in I\cap \hat{I}} (x_1 - \eta_n x_2)\,, 
\qquad
t_1= - b(x_1, x_2) \prod_{n\in D\setminus \{I\cup \hat{I}\}} 
(x_1 - \eta_n x_2)\,. 
\ee
At this point, $b(x_1,x_2)$ is an arbitrary polynomial that will get  
constrained by demanding that this solution is not BRST exact. BRST
exact solutions are of the form
$t_0=\hat{J} \phi_0- \phi_1 J$, so that 
\be
b(x_1,x_2) \in \frac{\BC[x_1,x_2]}{\langle s_1, s_2\rangle}
\,, 
\ee
where
\be
s_1=\prod_{n\in I\setminus \{ I \cap \hat{I} \} } (x_1 - \eta_n x_2 )\,, 
\qquad
s_2 =  \prod_{n\in \hat{I} 
\setminus \{ I \cap \hat{I}\} } (x_1 - \eta_n x_2)\,.
\ee
Since $s_1$ and $s_2$ do not have common factors, the dimension of
this ring is 
\be
\label{fermtop}
\deg s_1 \cdot \deg s_2= 
|I\setminus \{ I \cap \hat{I} \}|\cdot
|\hat{I} \setminus \{ I \cap \hat{I}\}|\,. 
\ee
This concludes the counting of the fermions. Note that this is just
the number of intersections of the set of lines 
$$
\bigcup_{n \in I\setminus \{ I\cap \hat{I} \}} \{ x_1 - \eta_n x_2 \} 
$$ 
with the set of lines 
$$
\bigcup_{n \in \hat{I}\setminus \{ I\cap \hat{I} \}} 
\{ x_1 - \eta_n x_2 \}\,,
$$
which is what the geometrical picture predicts.
\smallskip

\noindent Turning to the bosonic spectrum, we divide the first BRST
invariance condition 
\ba\nonumber
0 &=& \hat{J}\, \phi_0 - \phi_1\,  J \\
0 &=& \hat{E}\, \phi_1 - \phi_0 \, E 
\ea
by the greatest common divisor of $J$ and $\hat{J}$. Then we
find that 
\be
\phi_1 = a(x_1,x_2) \prod_{n\in \hat{I}\setminus \{I\cap \hat{I}\}} 
(x_1 - \eta_n x_2), \qquad
\phi_0= a(x_1, x_2) \prod_{n\in I\setminus \{I\cap \hat{I}\}} 
(x_1 - \eta_n x_2) \,. 
\ee
For this not to be BRST exact, the polynomial $a$ has to be in the
quotient ring of $\BC[x_1,x_2]$ by the ideal generated by $s'_1$ and 
$s'_2$, where 
\be
s'_1=\prod_{n\in I \cap \hat{I} } (x_1 - \eta_n x_2 )\,, \qquad
s'_2 =  \prod_{n\in D \setminus \{ I \cup \hat{I}\} } 
(x_1 - \eta_n x_2)
\,.
\ee
This means that the number of bosons equals 
\be\label{mppfin}
|{\rm bosons}| =\deg s'_1 \deg s'_2 = 
|I \cap \hat{I}|| D \setminus \{ I \cup \hat{I}\}|
\,.
\ee

\subsection{The spectrum between a tensor product and a rank $1$
brane}

To compute the spectrum between the tensor product brane $Q$ and the
rank $1$ brane $\hat{Q}$, we have to do again essentially the same
computation as before for the case of a single minimal model, except
that now the morphisms 
$(\phi_0,\phi_1)$ are two-component row vectors, 
$\phi_0 = (\phi_0^1,\phi_0^2)$ and 
$\phi_1 = (\phi_1^1,\phi_1^2)$. Likewise the fermions $(t_0,t_1)$ have
now also two components, $t_0=(t_0^1, t_0^2)$ and 
$t_1=(t_1^1,t_1^2)$. We will restrict ourselves to the case of a
permutation brane consisting of a single line ($\hat{J}$  linear), and
an arbitrary tensor product brane labelled by
$(\ell_1=L_1+1,\ell_2=L_2+1)$. (The case $\ell_1=\ell_2=1$ was
treated in \cite{ADD}.) 

\noindent The BRST conditions for the bosons are 
\ba 
\phi_1^1 x_2^{\ell_2} + \phi_1^2 x_1^{d-\ell_1} - 
\hat{J} \phi_0^1 &=& 0 \\
\phi_1^1 x_1^{\ell_1} - \phi_1^2 x_2^{d-\ell_2} - 
\hat{J} \phi_0^2 &=& 0 \\
\phi_0^1 x_2^{d-\ell_2} + \phi_0^2 x_1^{d-\ell_1} - 
\hat{E} \phi_1^1 &=& 0 \\ 
\phi_0^1 x_1^{\ell_1} - \phi_0^2 x_2^{\ell_2} - 
\hat{E} \phi_1^2 &=& 0 \,.
\ea
We can formally solve the first two equations by
\ba\label{formal}
\phi_0^1 &=& \frac{1}{\hat{J}} \left(\phi_1^1 x_2^{\ell_2} + 
\phi_1^2 x_1^{d-\ell_1}\right)\\
\phi_0^2 &=& \frac{1}{\hat{J}} 
\left(\phi_1^1 x_1^{\ell_1} - \phi_1^2 x_2^{d-\ell_2} \right) \,.
\nonumber 
\ea 
It is easy to see that this solution also solves the third and fourth
equation. We now have to discuss under which conditions the
expressions (\ref{formal}) are polynomials. This will be the case
whenever the zeros of the polynomial in the numerator contain those of
the denominator, which translates to the following relation between
$\phi_1^1$ and $\phi_1^2$
\be\label{choice}
\phi_1^1 = - \phi_1^2 \eta^{d-\ell_1-k} x_1^k x_2^{d-\ell_1-\ell_2-k}\,,
\ee
where $\eta$ is the root that appears in $\hat{J}=x_1-\eta x_2$, and
$k$ is an arbitrary integer. In particular, we therefore see 
that for a given $\phi_1^2$ there are a number of choices for
$\phi_1^1$. Once $\phi_1^1$ and $\phi_1^2$ are chosen, the remaining
components $\phi_0^1$ and $\phi_0^2$ are determined uniquely by
(\ref{formal}). 

\noindent These bosonic degrees of freedom are BRST trivial if 
\ba
\phi_1^1 &=&t_0^1 x_2^{d-\ell_2} + t_0^2 x_1^{d-\ell_1} 
+ (x_1-\eta x_2) t_1^1 \\
\phi_1^2 &=&t_0^1 x_1^{\ell_1} - t_0^2 x_2^{\ell_2} 
+ (x_1-\eta x_2) t_1^2 \,.\label{bostriv}
\ea
It is convenient to introduce the new coordinates $z=x_1-\eta x_2$ and
$x_2$. Choosing different polynomials $t_0^1,t_0^2$, we can eliminate 
the $z$-dependence of $\phi_1^1$ and $\phi_1^2$. As a consequence,
$\phi_1^1$ and $\phi_1^2$ are effectively polynomials of one variable
only. In particular, the choices of $\phi_1^1$ for a given $\phi_1^2$
in (\ref{choice})  are all equivalent, such that all components 
$\phi_0^1,\phi_0^2,\phi_1^1$ are uniquely determined by
$\phi_1^2$. The possible choices for $\phi_1^2$ are constrained by
(\ref{bostriv}), from which it follows that 
\be\label{pertenre}
\phi_1^2 \in \BC[x_2]/\langle x_2^{\ell_{min}}\rangle\,,
\quad {\rm where} \quad
\ell_{min} = \min \{ \ell_1, \ell_2 \}  \,.
\ee

\noindent The fermionic spectrum can be determined similarly. The BRST
conditions are    
\ba 
t_1^1 x_2^{\ell_2} + t_1^2 x_1^{d-\ell_1} + \hat{E} t_0^1 &=& 0 \\
t_1^1 x_1^{\ell_1} - t_1^2 x_2^{d-\ell_2} + \hat{E} t_0^2 &=& 0 \\
t_0^1 x_2^{d-\ell_2} + t_0^2 x_1^{d-\ell_1} + \hat{J} t_1^1 &=& 0 \\
t_0^1 x_1^{\ell_1} - t_0^2 x_2^{\ell_2} + \hat{J} t_1^2 &=& 0\,.
\ea
In analogy to the discussion of the bosons, one solves the last two 
equations for $t_1^1$ and $t_1^2$ and shows that the first two
equations are then automatically satisfied. From the requirement that
the formal solutions are in fact polynomials we derive 
\be
t_0^1 = -\eta^{d-\ell_1-k} x_2^{\ell_2-\ell_1-k} x_1^k t_0^2\,,
\ee
where we have made the additional assumption that 
$\ell_1\leq \ell_2$, and used that $\eta^d=-1$. The operator is BRST
exact if 
\ba\label{fermtriv}
t_0^1 &=&- \phi_1^1 x_2^{\ell_2} - \phi_1^2 x_1^{d-\ell_1} 
+ (x_1-\eta x_2) \phi_0^1 \\
t_0^2 &=& - \phi_1^1 x_1^{\ell_1} + \phi_1^2 x_2^{d-\ell_2} 
+ (x_1-\eta x_2) \phi_0^2 \,.
\ea
Choosing coordinates $z,x_2$ as above, it can be seen that the
$z$-dependence of $t_0^1$ and $t_0^2$ can be eliminated. The highest
power of $x_2$ for $t_0^2$ is $\ell_1-1$; since we have assumed that
$\ell_1\leq \ell_2$ this equals $\min \{\ell_1, \ell_2 \} -1$. Hence,
the number of fermions propagating between the two branes is 
$\min \{\ell_1, \ell_2 \}$ and equals the number  
of bosons. This was to be expected since the tensor product branes are
uncharged and therefore the Witten index with any other brane is
zero.

\section{Twisted NS-representations}

As a simple example we consider the case of the tensor product of two
$N=2$ minimal models with $k=1$. For $k=1$ (for which $c=1$) the
conformal weights of the NS representations are $h=0$ (for $[0,0,0]$)
and $h=1/6$ (for $[1,\pm 1,0]$). If the characters that appeared in
the overlap between the permutation and the tensor product brane were 
NS-characters evaluated at $\tilde{q}^{1/2}$, then we would have 
\begin{eqnarray}
\chi_{[0,0,0]}(\tilde{q}^{1/2}) & = &  \tilde{q}^{-1/48} 
\left(1 + {\cal O}(\tilde{q}^{1/2}) \right) 
= \tilde{q}^{h^d-\frac{1}{12}} 
\left(1 + {\cal O}(\tilde{q}^{1/2}) \right)\,,  \;\;
h^d = \frac{1}{16} 
\label{app1} \\
\chi_{[1,\pm 1,0]}(\tilde{q}^{1/2}) & = &  \tilde{q}^{1/16} 
\left(1 + {\cal O}(\tilde{q}^{1/2}) \right) 
= \tilde{q}^{h^d-\frac{1}{12}} 
\left(1 + {\cal O}(\tilde{q}^{1/2}) \right)\,,  \quad
h^d = \frac{7}{48} 
\,. \label{app2}
\end{eqnarray}
These characters now have to be interpreted as NS-characters of the
diagonal $N=2$ algebra, whose central charge is $c=2$. This is again a
minimal model (with $k=4$), and therefore the allowed conformal
weights are known. One easily checks that neither $h^d=1/16$ nor
$h^d=7/48$ are among the allowed list of conformal weights. Thus the
above characters {\it cannot} be interpreted in terms of
NS-representations of the diagonal $N=2$ algebra. 

The situation is different if the characters that appear in
the overlap between the permutation and the tensor product brane are 
R-characters (as we have argued they must). The conformal weights of
the R-representations for $k=1$ are $h=1/24$ (for $[0,\pm 1,1]$) and 
$h=3/8$ (for $[1,0,1]$). Instead of (\ref{app1}) and (\ref{app2}) we
then have 
\begin{eqnarray}
\chi_{[0,\pm 1,1]}(\tilde{q}^{1/2}) & = &  \tilde{q}^{0} 
\left(1 + {\cal O}(\tilde{q}^{1/2}) \right) 
= \tilde{q}^{h^d-\frac{1}{12}} 
\left(1 + {\cal O}(\tilde{q}^{1/2}) \right)  \quad\quad
h^d = \frac{1}{12} 
\,,  \\
\chi_{[1,0,1]}(\tilde{q}^{1/2}) & = &  \tilde{q}^{1/6} 
\left(1 + {\cal O}(\tilde{q}^{1/2}) \right) 
= \tilde{q}^{h^d-\frac{1}{12}} 
\left(1 + {\cal O}(\tilde{q}^{1/2}) \right)  \quad\;
h^d = \frac{1}{4} 
\,. 
\end{eqnarray}
The corresponding diagonal weights are then indeed allowed conformal
weights for the $k=4$ theory: $h=1/12$ is the conformal weight
of the NS-representation $[1,\pm 1,0]$, while $h=1/4$ is the conformal
weight of the NS-representation $[3,\pm 3,0]$.

\section{Gepner construction}

As is explained in the main part of the paper, the boundary states for
the Gepner model can be made by tensoring together constituent
boundary states of the branes in the $N=2$ minimal model before GSO
projection, followed by an orbifold projection.
For a single minimal model, the B-type boundary state
is 
\be
|\!|L,S\rangle\!\rangle_{(-1)^{(s+1)F}} = 
(2k+4)^{\frac{1}{4}} e^{-\pi i \frac{Ss}{2}}
\sum_{l}\, \sum_{\nu\in\Zop_2} 
\frac{S_{Ll}}{\sqrt{S_{0l}}} (-1)^{S\nu}
|[l,0,s+2\nu]\rangle\!\rangle \,
\ee
where $s=0$ for the NS-NS and $s=1$ for the R-R sector, and the sum
runs over all $l$ such that $l+s$ is even. In the above
notation, the subscript refers to the periodicity conditions of the
closed string fields on the circle; thus $s=0$ corresponds to the
NS-NS sector, while $s=1$ is the R-R sector. Note that we can recover
the boundary states constructed in section~2 by adding a NS-NS
boundary state of the unprojected theory and a R-R boundary state with
the appropriate normalisation factor of $1/\sqrt{2}$. 

The boundary state is invariant under the $\ZZ_{k+2}$ axial symmetry 
of the minimal model. 
For the Gepner construction we  need to formulate the boundary states
also in the sector twisted by $g^n$, where $g$ is the generator of
the axial symmetry
\be
|\!|L,\hat{M},S\rangle\!\rangle_{(-1)^{(s+1)F}g^n} = 
(2k+4)^{\frac{1}{4}} e^{-\pi i \frac{\hat{M}n}{k+2} -\pi i \frac{Ss}{2}}
\sum_{l} \sum_{\nu\in \Zop_2} \frac{S_{Ll}}{\sqrt{S_{0l}}} (-1)^{S\nu}
|l,n,s+2\nu \rangle\!\rangle \,.
\ee
Here $|l,n,s+2\nu \rangle\!\rangle$ is the Ishibashi state in the
sector $\H_{[l,n,s+2\nu]}\otimes \bar\H_{[l,-n,-s-2\nu]}$, and
the sum runs only over those values of $l$ for which $l+n+s$ is even. 
In the  open string sector, the twist leads to an insertion of
$g^n$. It is then clear that we need a further label that specifies
the action of the global symmetry on the Chan-Paton labels. This is
the origin of the label $\hat{M}$ that appears only in the overall
phase of the twisted boundary state. In order for this formula to make
sense, we need that $L+\hat{M}+S$ is even. 

The one-loop overlap between two such states is then (in the sector
labelled by $s$ and $n$)
\begin{eqnarray}
\langle\!\langle L',\hat{M}',S'& |\!| &
q^{\frac{1}{2}(L_0+\bar{L}_0)-\frac{c}{24}}
|\!| L,\hat{M},S \rangle\!\rangle_{(-1)^{(s+1)F}g^n} 
\nonumber \\
& & = \sum_{[l',m',s']} 
\delta^{(2)}(S+S'+s')\, 
e^{i\pi \frac{n}{k+2} (\hat{M}'-\hat{M} + m') } \,
e^{i\pi \frac{s}{2} (S'-S-s')} \nonumber \\
& & \qquad \qquad \times 
\left( N_{L'L}^{l'} + (-1)^{n+s}\, N_{L' L}^{k-l'} \right)\,
\chi_{[l',m',s']}(\tilde{q}) \,.
\end{eqnarray}
Here the sum runs over all equivalence classes $[l',m',s']$. If we sum
over all triples $(l',m',s')$ such that $l'+m'+s'$ is even, we can
instead write the last expression as 
\begin{eqnarray}
&&\langle\!\langle L',\hat{M}',S' |\!| 
q^{\frac{1}{2}(L_0+\bar{L}_0)-\frac{c}{24}}
|\!| L,\hat{M},S \rangle\!\rangle_{(-1)^{(s+1)F}g^n}  \label{Ct}\\
& & \quad = \quad \sum_{(l',m',s')} 
\delta^{(2)}(S+S'+s')\, 
e^{i\pi \frac{n}{k+2} (\hat{M}'-\hat{M} + m') } \,
e^{i\pi \frac{s}{2} (S'-S-s')}\,
N_{L' L}^{l'}\, \chi_{[l',m',s']}(\tilde{q}) \,. \nonumber
\end{eqnarray}
The tensor product boundary states of \cite{RS} can in this notation
be written as 
\begin{eqnarray}
& |\!| & L_1,\dots, L_5,\hat{M},S\rangle\!\rangle_{(-1)^{(s+1)F}}
= \frac{1}{\sqrt{H}}
\sum_{n \in \Zop_H} \bigotimes_{i=1}^5
|\!|L_i,\hat{M}_i,S_i\rangle\!\rangle_{(-1)^{(s+1)F}g^n} 
\nonumber \\
& & = \frac{\prod_i (2k_i+4)^{\frac{1}{4}}}{\sqrt{H}}
\sum_{n\in\Zop_H} \, 
e^{- \pi i \sum_i (\frac{\hat{M}_i n}{k_i+2} +\frac{S_i s}{2})} 
\bigotimes_{i=1}^5 \sum_{l_i} \sum_{\nu_i\in\Zop_2}
\frac{S_{L_il_i}}{\sqrt{S_{0l_i}}} (-1)^{S_i\nu_i}
|l_i,n,s+2\nu_i \rangle\!\rangle\,. \nonumber
\end{eqnarray}
Note that the boundary state on the left actually only depends on 
$\hat{M} = H \, \sum_i \frac{\hat{M}_i}{k_i+2}$.  
Furthermore, since the $S_i$ are either
all even or all odd (so that we choose the same sign for the gluing
condition $\eta$ in all five factors), the boundary state actually
only depends on $S=\sum_i S_i$.
\medskip

For the permutation boundary states the analysis is similar. Before
GSO-projection the permutation boundary state in the tensor product of
two minimal models is given by 
\begin{eqnarray}
|\!|L,M,S_1,S_2\rangle\!\rangle_{(-1)^{(s+1)F}} &=& \frac{1}{2}
\sum_{l,m}\, \sum_{\nu_1,\nu_2\in \Zop_2}  
\frac{S_{Ll}}{S_{0l}} e^{\pi i \frac{Mm}{k+2}}
\ (-1)^{S_1 \nu_1+S_2\nu_2} \ e^{-\pi i \frac{s}{2}(S_1+S_2)} \nonumber \\
&& \qquad \qquad \qquad |[l,m,s+2\nu_1] \otimes [l,-m ,s+2\nu_2]
\rangle\!\rangle \,.
\end{eqnarray}
Here $L+M$ and $S_1+S_2$ are even, and the sum runs over all $l,m$
such that $l+m+s$ is even. As before this boundary state is invariant
under $g_1\, g_2$, but not under $g_1$ or $g_2$ individually, which
shift the $M$-label by $\pm 2$. To construct the permutation boundary
states in the sectors twisted by $g=g_1 g_2$, we
observe that the permutation gluing condition requires that 
$\bar{m}_2 = -m_1$ and $m_2 = - \bar{m}_1$. In the sector twisted by
$g^n$ the relation between left and right-moving $m$-labels is
$m_1=\bar{m}_1 +2n$, $m_2=\bar{m}_2 +2n$, so that the relevant 
Ishibashi states have labels $m_2=-\bar{m}_1 = -m_1+2n$. Therefore,
the boundary state takes the form
\begin{eqnarray}
|\!|L,M,\hat{M},S_1,S_2\rangle\!\rangle_{(-1)^{(s+1)F}g^n} & = & 
\frac{1}{2} e^{-\frac{\pi i n}{k+2}(M+\hat{M})}
\sum_{l,m} \sum_{\nu_,\nu_2\in \Zop_2}  
\frac{S_{Ll}}{S_{0l}}\, e^{\pi i \frac{Mm}{k+2}} \,
(-1)^{S_1 \nu_1+S_2\nu_2} \nonumber \\
&& \; e^{-\pi i \frac{s}{2}(S_1+S_2)} \, 
|[l,m,s+2\nu_1]\otimes [l,-m+2n ,s+2\nu_2]\rangle\!\rangle
\,, \nonumber 
\end{eqnarray}
where, as before, an additional label $\hat{M}$ had to be introduced.
We require that $M+\hat{M}$ is always even, so that the boundary state
is invariant under $n\to n+k+2$. Also, as before in the discussion of
section~4, $L+M$ and $S_1+S_2$ are even.

With this ansatz we then obtain the following one-loop amplitudes
between twisted permutation boundary states (in the sector labelled by
$s$ and $n$)
\begin{eqnarray}
&&\langle\!\langle L',M',\hat{M}',S_1',S_2'|\!|
q^{\frac{1}{2}(L_0+\bar{L}_0)-\frac{c}{12}}
|\!| L,M,\hat{M},S_1,S_2 \rangle\!\rangle_{(-1)^{(s+1)F}g^n} 
= 
\sum_{[l_1,m_1,s_1] , [l_2,m_2,s_2]} 
\nonumber \\
& & \quad
e^{-\frac{\pi i s}{2}(S_1-S_1'+s_1+S_2-S_2'+s_2)} \,
\delta^{(2)}({S_1-S_1'+s_1})\, \delta^{(2)}({S_2-S_2'+s_2}) \,
e^{\frac{\pi i n}{k+2}(m_1+m_2-\hat{M} +\hat{M}')} \nonumber \\ 
&&\qquad 
\sum_l\, \left( N_{LL'}^l N_{l_1l_2}^l \delta^{(2k+4)}({M-M'+m_1-m_2})
\right. \nonumber \\
& & \qquad \qquad \left. 
+(-1)^{n+s} N_{LL'}^lN_{l_1k-l_2}^l \delta^{(2k+4)}({M-M'+m_1-m_2+k+2})
\right) \nonumber \\
&&\qquad \qquad \qquad \qquad \qquad \qquad 
\chi_{[l_1,m_1,s_1]}(\tilde{q})\,
\chi_{[l_2,m_2,s_2]}(\tilde{q}) \,.
\end{eqnarray}
In the above equation, the sum runs again over equivalence classes 
$[l_i,m_i,s_i]$. If we relax this condition and sum over all triples
with $l_i+m_i+s_i$ even, we obtain instead
\begin{eqnarray}
&&\langle\!\langle L',M',\hat{M}',S_1',S_2'|\!|
q^{\frac{1}{2}(L_0+\bar{L}_0)-\frac{c}{12}}
|\!| L,M,\hat{M},S_1,S_2 \rangle\!\rangle_{(-1)^{(s+1)F}g^n} 
= \half 
\sum_{(l_1,m_1,s_1) , (l_2,m_2,s_2)} 
\nonumber \\
& & \quad
e^{-\frac{\pi i s}{2}(S_1-S_1'+s_1+S_2-S_2'+s_2)} \,
\delta^{(2)}({S_1-S_1'+s_1})\, \delta^{(2)}({S_2-S_2'+s_2}) \,
e^{\frac{\pi i n}{k+2}(m_1+m_2-\hat{M} +\hat{M}')} \nonumber \\ 
&&\qquad 
\sum_l\, N_{LL'}^l N_{l_1l_2}^l \delta^{(2k+4)}({M-M'+m_1-m_2})
\, \chi_{[l_1,m_1,s_1]}(\tilde{q})\,
\chi_{[l_2,m_2,s_2]}(\tilde{q}) \,. \label{Cp}
\end{eqnarray}
Since $M+\hat{M}$ is even it is immediate (from the $\delta^{(2k+4)}$
constraint) that $m_1+m_2 - \hat{M} - \hat{M}'$ is even. Furthermore,
the form of the amplitude shows explicitly that the twist leads to an
insertion of the group action in the open string sector, where the
action on the Chan-Paton labels is specified by $\hat{M}$ and
$\hat{M}'$, respectively. 

The overlap between a permutation and a tensor product boundary state
(in two factors) is again subtle; taking into account the phase of 
(\ref{import1}) one finds 
\begin{eqnarray}
&&\langle\!\langle L_1,L_2,\hat{M}',S_1',S_2'|\!|
q^{L_0+\bar{L}_0 - \frac{c}{24}} |\!| L,M,\hat{M}, S_1,S_2
\rangle\!\rangle_{(-1)^{F(s+1)}g^n}  \nonumber \\
& & \qquad \qquad = \sum_{(l',m',s')}  \,
\chi_{[l',m',s']}(\tilde{q}^{\frac{1}{2}}) \,
e^{\frac{\pi i n}{k+2}(\hat{M}'-\hat{M}+m'+ 1)}
e^{-\frac{\pi i s}{2}(S_1+S_2-S_1'-S_2'+s'+1)} \\
& & \qquad \qquad \qquad \delta^{(2)}({S_1-S_1'+S_2-S_2'+s'+1}) \,
\sum_{\hat{l}} N_{L_1L_2}^{\hat{l}} \, N_{Ll'}^{\hat{l}}  \,.
\nonumber
\end{eqnarray}
Here the sum runs again over the triples $(l',m',s')$ such
that $l'+m'+s'$ is even. One can easily check that
$\hat{M}'-\hat{M}+m'+1$ is always even. Furthermore, as before in the
untwisted case (that was discussed in section~4), $s'$ is always
odd if the two spin structures of the boundary states are the same. In
this case, the representations with $s'$ odd should be interpreted as
twisted NS sector representations.

\end{document}